\documentclass[useAMS,usegraphicx]{mn2e}
\usepackage{times}

\newif\ifAMStwofonts
\AMStwofontstrue

\newcommand{\me}{\mathrm{e}}
\newcommand{\mpi}{\mathrm{\pi}}
\newcommand{\mi}{\mathrm{i}}

\newcommand{\et}{et al.\ }

\newcommand{\xmm}{{\it XMM-Newton}}

\newcommand{\ls}
{\mathrel{\hbox{\rlap{\hbox{\lower4pt\hbox{$\sim$}}}\hbox{$<$}}}}
\newcommand{\gs}
{\mathrel{\hbox{\rlap{\hbox{\lower4pt\hbox{$\sim$}}}\hbox{$>$}}}}

\def\mrk{{Mrk 766}}

\title[Properties of X-ray light curves from AGN]
{On characterising the variability properties of X-ray light curves from active galaxies}

\author[Vaughan \et]
       {S. Vaughan,$^{1,2}$
        R. Edelson,$^{3}$
        R. S. Warwick$^{2}$ and
        P. Uttley$^{4}$ \\
$^{1}$Institute of Astronomy, Madingley Road, Cambridge CB3 0HA\\
$^{2}$X-Ray and Observational Astronomy Group, Department of Physics and Astronomy, University of Leicester, Leicester LE1 7RH\\
$^{3}$Astronomy Department, University of California, Los Angeles, CA 90095-1562; USA \\
$^{4}$Department of Physics and Astronomy, University of Southampton, Southampton SO17 1BJ }

\date{Accepted: 24/7/2003; Submitted: 23/7/2003; in original form: 3/3/2003}
\pagerange{\pageref{firstpage}--\pageref{lastpage}}
\pubyear{2003}

\begin{document}
\maketitle
\label{firstpage}

\begin{abstract}
We review some practical aspects of measuring the amplitude of
variability in `red noise'  light curves typical of those from Active
Galactic Nuclei (AGN). The quantities commonly used to estimate the
variability amplitude in AGN light curves, such as the fractional rms
variability amplitude, $F_{\rm{var}}$, and excess variance,
$\sigma_{\rm{XS}}^{2}$, are examined. Their statistical properties,
relationship to the power spectrum, and uses for investigating the
nature of the variability processes are discussed.  We demonstrate
that $\sigma_{\rm{XS}}^{2}$ (or similarly $F_{\rm{var}}$) shows large
changes from one part of the light curve to the next, even when the
variability is produced by a stationary process. This limits the
usefulness of these estimators for quantifying differences in
variability amplitude between different sources or from epoch to epoch
in one source.  
Some examples of the expected scatter in the variance are 
tabulated for various typical power spectral shapes, based
on Monte Carlo simulations. The excess variance can be useful for
comparing the variability amplitudes of light curves in different
energy bands from  the same observation. Monte Carlo simulations are
used to derive a  description of the uncertainty in the amplitude
expected between different energy bands (due to measurement errors).
Finally, these estimators are used to demonstrate some variability
properties of the bright Seyfert 1 galaxy Markarian 766.  The source
is found to show a strong, linear correlation between rms amplitude
and flux, and to show  significant spectral variability.
\end{abstract}


\section{Introduction}

One of the defining characteristics of Active
Galactic Nuclei (AGN) is that their X-ray emission is variable. X-ray
light curves from 
Seyfert 1 galaxies show unpredictable and seemingly aperiodic
variability (Lawrence \et 1987; M$^{\rm c}$Hardy 1989). 
Such random variability is often referred to as {\it noise},
meaning that it is the result of a stochastic, as opposed to
deterministic, process. In this context the `noise' is intrinsic
to the source and not a result of measurement errors (such as Poisson
noise), i.e. the signal itself is the output of a noise process. 

One of the most common tools for examining AGN variability (and noise
processes in general) is the fluctuation Power Spectral Density (PSD)
which represents the amount of variability power (mean of the squared
amplitude) as a function of temporal frequency (timescale$^{-1}$).
The high frequency PSDs of Seyferts are usually well-represented by
power-laws over 
a broad range of frequencies ($\mathcal{P}(f) \propto f^{-\alpha}$,
where $\mathcal{P}(f)$ is the power at frequency $f$) with slopes
$\alpha=1-2$ (Green, M$^{\rm c}$Hardy \& Lehto, 1993; Lawrence \&
Papadakis, 1993; Edelson \& Nandra, 1999; Uttley, M$^{\rm c}$Hardy \&
Papadakis, 2002; Vaughan, Fabian \& Nandra 2003; Markowitz \et
2003). Such a spectrum, with a slope $\alpha \gs 1$ is usually called
`red noise' (for an introduction to red noise see Press 1978).

If Seyfert~1 light curves are viewed as the product of a stochastic
(in this case red noise) process then the specific details of each
individual light curve provide little physical insight. Each light
curve is only one {\it realisation} of the underlying stochastic
process, i.e. it is one of the {\it ensemble} of  random light curves
that might be generated by the process. Each new realisation will
look different and these changes are simply statistical fluctuations
inherent in any stochastic process (as opposed
to changes in the nature of the process itself). 
Therefore one should expect two light curves to have different
characteristics (such as mean and variance) even if they are
realisations of the same process.  
On the other hand, data from deterministic processes, for example the
energy spectrum of
a non-varying source or the light curve of a strictly periodic source
(such as a pulsar), should be repeatable within the limits set
by the measurement errors.

It is the average properties of the variability (such as the PSD) that
often provide most insight into the driving process.  For example, the
PSD of any real red noise process cannot continue as a steep power-law
indefinitely to longer timescales or the integrated variability power
would diverge. Therefore the PSDs of AGN variability must break to a
flatter index at low frequencies; the position of such a break would
represent a characteristic variability timescale and may yield
information about the underlying driving process.  Recent timing
studies have indeed found evidence that the steep power-law PSDs of
Seyfert~1s show a flattening, or turnover, at low frequencies (Edelson
\& Nandra 1999; Uttley \et 2002; Markowitz \et 2003).

In many cases however the data are not adequate for PSD analysis.  In
these situations the variability is usually described in terms of the
statistical moments (e.g. the sample mean and variance, etc.). However, due to
the stochastic nature of red noise variability there is a large degree
of randomness associated with these quantities.   In
practice this means that it is difficult to assign meaningful
errors to the variance. This in turn makes it difficult to
quantitatively compare variances, say from repeated observations of
the same source (and thereby test whether the variability is {\it
stationary}). 
Such an analysis might be desirable; it could in
principle reveal changes in the `state' of the source if its
variability properties were found to evolve with time.  This paper
discusses this and related problems that are encountered when
examining the variability properties of AGN.  Particular emphasis is
placed on the mathematical properties and implications of the inherent
randomness in the variability. The mathematical details are well
understood from the general theory of stochastic processes (e.g.
Priestley 1981 for spectral analysis) but some of the practical
consequences for AGN observations have not been discussed in detail.
On the basis of simulated data some recipes are developed that may
serve as a useful guide for observers wishing make quantitative use of
their variability analysis without the recourse to e.g. extensive Monte
Carlo simulations.

The paper is organised as
follows.  Section~\ref{sect:stats} defines the estimators to be
discussed (namely the periodogram and the variance).   
Simulated data are used to illustrate various aspects of
these estimators; section~\ref{sect:simulations} describes methods for
producing artificial red noise time series.
Section~\ref{sect:stationarity} discusses the stationarity of time
series. 
Sections~\ref{sect:errors_intrinsic} and \ref{sect:errors_poisson}
discuss two sources of  uncertainty associated with 
measuring variability amplitudes, the first due to the stochastic nature of the
variability and the second due to flux measurement errors. 
Section~\ref{sect:mrk766} gives an example using a real \xmm\
observation of \mrk.   
Finally, a brief discussion of these results is given in
section~\ref{sect:disco} and the conclusions are summarised in
section~\ref{sect:conc}. 

\section{Estimating the Power Spectral Density}
\label{sect:stats}

The PSD defines the amount of variability `power' as a function of temporal
frequency. It is estimated by calculating the {\it
  periodogram}\footnote{ Following Priestley 
(1981) the term ``periodogram'' is used for the discrete function
$P(f_{j})$, which is an estimator of the continuous PSD $\mathcal{P}(f)$. The
periodogram is therefore specific to each realisation of the
process, whereas the PSD is representative of the true, underlying
process.} 
(Priestley 1981; Bloomfield 2000). 

For an evenly sampled light curve (with a sampling
period $\Delta T$) the periodogram is the modulus-squared of the
Discrete Fourier Transform 
(DFT) of the data (Press \et 1996). For a light curve comprising a series
of fluxes $x_i$ measured at discrete times $t_{i} ~ (i=1,2,\ldots,N)$:
\begin{eqnarray}
\label{eqn:ft}
\lefteqn{
|DFT(f_{j})|^{2} =
\left| \sum_{i=1}^{N} x_i~\me^{2\mpi \mi f_{j} t_{i}} \right|^{2} = }
\nonumber\\ 
\lefteqn{ \qquad
\left\{ \sum_{i=1}^{N} x_i~\cos (2 \mpi f_{j} t_i) \right\}^{2}
+ \left\{ \sum_{i=1}^{N} x_i~\sin (2 \mpi f_{j} t_i) \right\}^{2},
}
\end{eqnarray}
at $N/2$ evenly spaced frequencies  $f_{j}=j/N \Delta T$ (where $j=1, 2,
\ldots, N/2$), $f_{N/2}=1/2\Delta T$ is the  Nyquist frequency,
$f_{\rm Nyq}$.  Note that it  is 
customary to subtract the mean flux from the light curve before
calculating the DFT. This eliminates the zero-frequency power.  The
periodogram, $P(f_{j})$, is then calculated by choosing an appropriate
normalisation $A$ (see Appendix~\ref{sect:pds_norm} for more on
periodogram normalisations). For example
\begin{equation}
\label{eqn:pds}
P(f_{j}) = A |DFT(f_{j})|^{2} = \frac{2\Delta T}{N} |DFT(f_{j})|^{2}.
\end{equation}

If the time series is a photon counting signal such as normally
encountered in X-ray astronomy, and is binned into intervals of
$\Delta T$, the effect of Poisson noise is to add an
approximately constant amount of power to the periodogram at all
frequencies. With the above normalisation this constant Poisson noise
level is $2\bar{x}$ (assuming the light curve is not background
subtracted). 

\subsection{Statistical properties of the periodogram}

\begin{figure}
 \begin{center}
  \rotatebox{-90}{
   \includegraphics[width=6 cm]{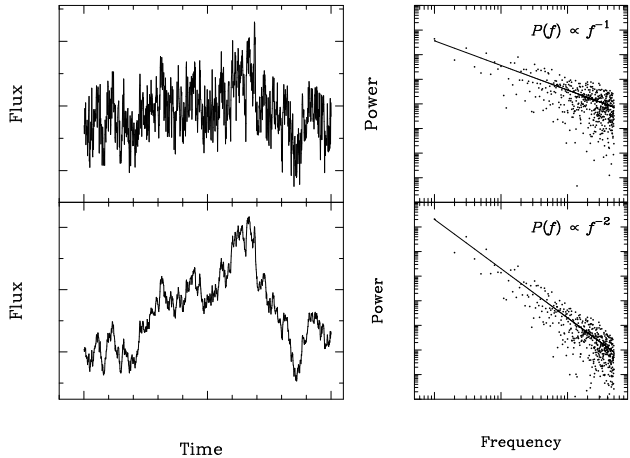}
  }
  \caption{
Simulated time series (left) and their periodograms (right).
The upper panel shows a `flicker noise' time series which has a $f^{-1}$
PSD. The lower panel shows a `random walk' time series with a
$f^{-2}$ PSD.
Note the large scatter in the periodogram (dots) around the underlying PSD (solid line).
It is clear that the time series with the steeper PSD shows more
power in long-term variability while the time series with the
flatter PSD shows relatively more power in short term variability
(flickering).  
The two series were generated using the same random number sequence.
\label{fig:two_curves}
}
\end{center}
\end{figure}

The periodogram of a noise process, if measured from a single time
series, shows a 
great deal of scatter around the underlying PSD.  
In particular, the periodogram at a
given frequency [$P(f)$] is scattered around the 
PSD [$\mathcal{P}(f)$] following a $\chi^{2}$ distribution with two degrees of
freedom (van der Klis 1989):
\begin{equation}
\label{eqn:pds_scatter}
P(f) = \mathcal{P}(f) \chi_{2}^{2}/2,
\end{equation}
where $\chi_{2}^{2}$ is a random variable distributed as $\chi^{2}$
with two degrees of freedom, i.e. an exponential distribution with a
mean and variance of two and four, respectively.
The periodogram is distributed in this way 
because the real and imaginary parts of the DFT are normally
distributed for a stochastic process\footnote{
The DFT at the Nyquist frequency
is always real when $N$ is even so the periodogram at this frequency 
is distributed as $\chi_{1}^{2}$, i.e. with one degree
of freedom.} (section 6.2 of Priestley, 1981; Jenkins \& Watts 1968). 
The expectation value
of the periodogram is equal to the PSD but
its standard deviation is 100 per cent, leading to the larger scatter
in the periodogram (see Fig.~\ref{fig:two_curves}). See Leahy \et
(1983), van der Klis (1989), Papadakis \& Lawrence (1993), Timmer \&
K\"{o}nig (1995) and  Stella \et (1997) for further discussion of this
point. 

When applied to real data the periodogram is an {\it inconsistent}
estimator of the PSD, meaning that the scatter in the periodogram does
not decrease as the number of data points in the light curve increases
(Jenkins \& Watts 1968). In order to reduce this scatter the
periodogram must be smoothed (averaged) in some fashion.  As the
number of data points per bin increases (either by binning over
frequencies or averaging over many data segments) the scatter in the
binned periodogram decreases, i.e. the averaged periodogram is a {\it
consistent} estimator of the PSD (see Papadakis \& Lawrence 1993 and
van der Klis 1997 for more on binned periodogram estimates). A further
point is that  periodograms measured  from finite data tend to be
biased by windowing effects which further complicate their
interpretation (van der Klis 1989; Papadakis \& Lawrence 1993; Uttley
\et 2002 and see below).

\subsection{Integrated power}

The integral of the PSD between two frequencies ($f_{1}$ and
$f_{2}$) yields the contribution to the expectation value of the
(`true') variance due to 
variations between the corresponding
timescales ($1/f_{1}$ and $1/f_{2}$). This result follows from Parseval's
theorem (see e.g. van der Klis 1989)
\begin{equation}
\label{eqn:correspondence1}
\langle S^{2} \rangle = \int_{f_{1}}^{f_{2}} \mathcal{P}(f) df.
\end{equation}
Correspondingly, for a discrete time series the
integrated periodogram yields the 
observed variance for that particular realisation
\begin{equation}
\label{eqn:correspondence}
S^{2} 
= \sum_{j=1}^{N/2} P(f_{j}) \Delta f,
\end{equation}
where $\Delta f$ is the frequency resolution of the DFT ($\Delta
f=1/N\Delta T$). 
The total variance of a real light
curve is equal to its periodogram integrated over the frequency range
$f_{1}=1/N\Delta T$ to $f_{\rm Nyq}=1/2\Delta T$.

The sample variance (which will differ from observation to
observation) is given by:  
\begin{equation}
\label{eqn:variance}
S^{2} = \frac{1}{N-1} \sum_{i=1}^{N} (x_{i} - \bar{x})^{2},
\end{equation}
where $\bar{x}$ is the arithmetic mean of $x_{i}$.
In the limit of large $N$ these two variance estimates are identical.
The normalised variance\footnote{In 
AGN studies normalised quantities are often used in preference to 
absolute quantities as they are independent of the flux
of a specific source. This means that, in principle,
normalised amplitudes can be used to compare 
sources with different fluxes.}  is simply $S^{2}/\bar{x}^{2}$.

\section{Simulating red noise light curves}
\label{sect:simulations}

\subsection{Algorithms}

In order to elucidate the properties of the variance of red noise
data, random light curves were generated from
power-law PSDs similar to those of AGN.  Fig.~\ref{fig:two_curves}
shows two artificial time series and their periodograms. It is worth
reiterating that the large scatter in the periodograms is an intrinsic
property of stochastic processes -- it does not depend on the number
of data points and is not related to Poisson noise in the data.

These artificial time series were produced using the algorithm of
Timmer \& K\"{o}nig (1995).  This generates random time series with
arbitrary broad-band PSD, correctly accounting for the intrinsic
scatter in the powers (i.e.  equation~\ref{eqn:pds_scatter}).   Other
methods of generating random light curves include the related
`summing of sines' method (Done \et 1992). 
Note that it is not correct to randomise only the phases of the
component sine functions, their amplitudes must also be
randomised. Otherwise this method does not account for
this intrinsic scatter in the powers.  Shot-noise models can produce
red noise time series with certain PSD shapes (see Lehto 1989).  There
also exist various mathematical tricks for producing data with
specific power-law PSD slopes.  Data with a $\alpha=1$ PSD (often
called `flicker noise') can be generated using the half-integral
method outlined  in Press (1978), while $\alpha=2$ (`random walk')
data can be generated using a first-order autoregressive process
(AR[1]), essentially a running sum of Gaussian deviates (see Deeming
1970 and Scargle 1981 for more on such methods).  The method of Timmer
\& K\"{o}nig 
(1995) is used below as this can generate time series from an
arbitrary PSD and is computationally efficient.

\subsection{Simulating `realistic' data}

Some caution should be applied when using these routines to produce
artificial time series. As mentioned briefly in the previous section,
periodograms measured from real data tend to be biased by windowing
effects. For uninterrupted but finite observations data of red noise
processes the 
most important of these effects is `red noise leak' -- the transfer
of power from low to high frequencies by the lobes of the window
function (see e.g. Deeter \& Boynton 1982; van der Klis 1997). If
there is significant power at frequencies below the lowest frequency
probed by the periodogram (i.e. on timescales longer than the length
of the observation) this can give rise to slow rising or falling
trends across the light curve. These trends contribute to the variance
of the light curve. Thus variability power `leaks' into the data
from frequencies below the observed frequency band-pass.
The degree to which this occurs, and the resultant bias on the
measured periodogram, depend on the shape of the underlying PSD and
the length of the observation (Papadakis \& Lawrence 1995; Uttley \et
2002). For flat PSD slopes ($\alpha < 1.5$) the amount of leakage from
low frequencies is usually negligible.

Since AGN light curves usually  contain significant power on
timescales longer than those probed (see
section~\ref{sect:weak_non-stationarity})  the effects of red noise
leak must be included in simulations of AGN light curves.  This can be
achieved by using the Timmer \& K\"{o}nig (1995) algorithm to generate
a long light curve  from a PSD that extends to very low frequencies
and then using a light curve segment of the required length.  Data simulated in
such a fashion will include power on  timescales much longer than
covered in the short segment.
The effects of measurement errors (e.g. Poisson noise) can be
included in the simulations using standard techniques (e.g. Press \et
1996).

\section{Stationarity}
\label{sect:stationarity}

\begin{figure}
 \begin{center}
  \rotatebox{-90}{
   \includegraphics[width=8.5 cm, angle=90]{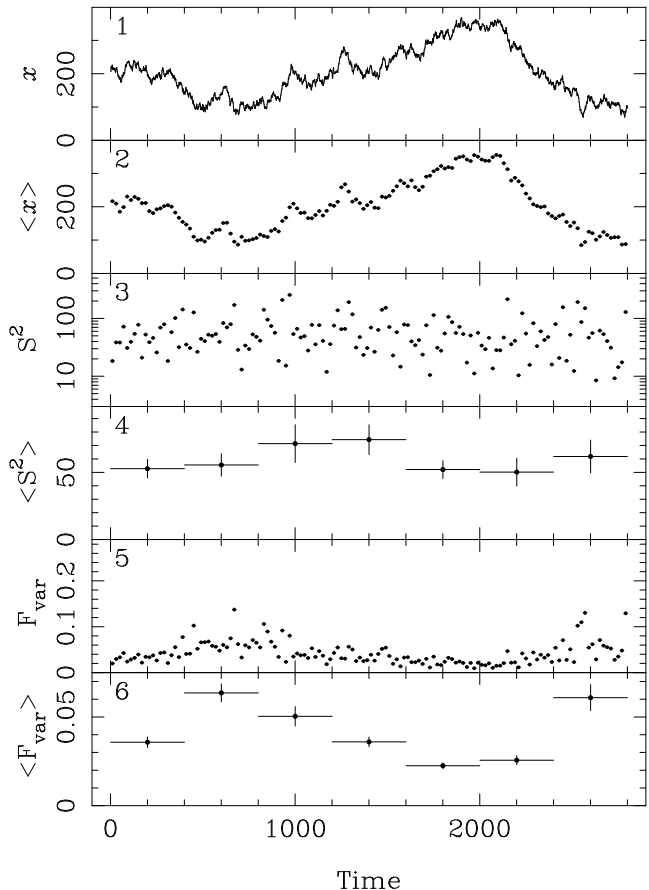}
  }
\caption{Panel 1: Simulated red noise time series (with a $f^{-2}$
PSD) with $N=2800$ points. 
Panel 2 and 3: mean and variance measured from segments of 20 points
(calculated using equation~\ref{eqn:variance}).
The variances follow a distribution of the form shown in the bottom
panel of Fig.~\ref{fig:fvar_dist1}.
(Note the variance is plotted on a logarithmic scale.)
Panel 4: averaged variance measured by binning
the individual variances into groups of 20 consecutive estimates.
The errors are the standard error on the mean (equation
4.14 of Bevington \& Robinson 1992).
These averaged variances are consistent with staying constant.
In other words, although the instantaneous value of the variance
fluctuates, its expectation value is consistent with constant
(i.e. a stationary process).
Panel 5: fractional rms amplitude ($\sqrt{S^2/\bar{x}^2}$) measured
from segments of 20 points. 
Panel 6: averaged fractional rms amplitude measured by binning
the individual amplitudes into groups of 20.
The fractional amplitude is anti-correlated with the light curve because
$\langle S^2 \rangle$ is constant but $F_{\rm var}$ is normalised by
the light curve flux.  
\label{fig:curve1}
}
\end{center}
\end{figure}

A stationary process is one for which the statistical properties (such
as mean, variance, etc.) do not depend on time. Fig.~\ref{fig:curve1}
shows an artificial red noise time series together with its mean and
variance measured every 20 data points.  The simulation was produced
from a single, well-defined PSD (which did not vary). The process is
therefore stationary. It would have been reasonable to expect the
resulting time series  (a realisation of the process) to appear
stationary. This is not the case however; both the mean and variance
change with time (panels 2 and 3). This has nothing whatsoever to do
with measurement errors - the simulation has zero errors. This
simulation demonstrates that, when dealing with red noise,
fluctuations in variance are not sufficient to claim the variability
process is non-stationary.

As the purpose of time series analysis is to gain insight into the
process, not the details of any specific realisation, a more robust
approach is needed to determine whether these data were produced by a
time-stationary process or a non-stationary process. It would be more
insightful to consider whether the {\it expectation values} of the
characteristics (such as the variance) are time-variable. The
expectation values should be representative of the properties of the
underlying process, not just any one realisation.  See section 1.3 of
Bendat \& Piersol (1986) for a discussion of this point.

\subsection{Weak non-stationarity}
\label{sect:weak_non-stationarity}

For a process with a steep red noise PSD ($\alpha \ge 1$), the
integrated periodogram will diverge as $f \rightarrow 0$.  This means
that (following equation~\ref{eqn:correspondence1}) the variance of a
red noise time series with a steep PSD will diverge with time.  In
this case there is no well-defined mean; Press \& Rybicki (1997)
describe this form of variability as `weakly non-stationary.'  For
time series with power spectra flatter than this the variance
converges as $f \rightarrow 0$, thus for a white noise process with a
flat PSD ($\alpha=0$), the variance will converge as the observation
length increases, and there will be a well-defined mean on long
timescales.

Of course, for any real process the PSD must eventually flatten such
that the power does not diverge (i.e. $\alpha < 1$ on sufficiently
long timescales).  
Thus, weak non-stationarity is entirely due to
observations  sampling only the steep part of the PSD of a source.
But in AGN this flattening occurs on timescales much longer than those
probed by typical observations.  
For instance, \xmm\ observations of AGN typically
last for $\sim {\rm few} \times 10^{4}$~s whereas   in many objects
the PSD is steep until $> 10^5$~s and in some cases probably much
longer (Edelson \& Nandra 1999; Uttley \et 2002; Markowitz \et 2003).
Therefore on the timescales relevant for most X-ray observations, AGN
light curves should be considered weakly non-stationary.

\subsection{Stochasticity}
\label{sect:stochasticity}

\begin{figure}
 \begin{center}
   \includegraphics[width=5.0 cm, angle=270]{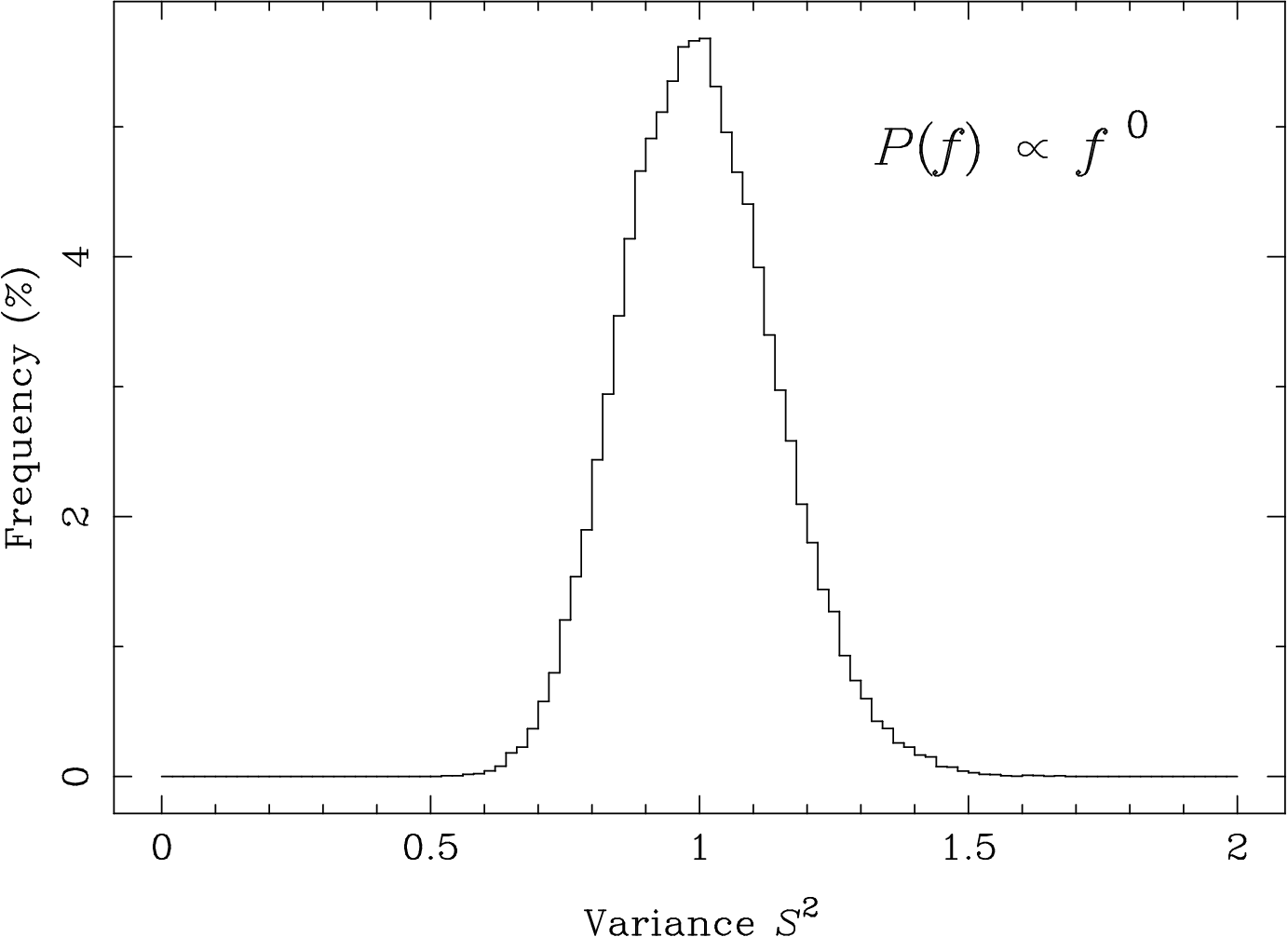}\\
\vspace{0.2 cm}
   \includegraphics[width=5.0 cm, angle=270]{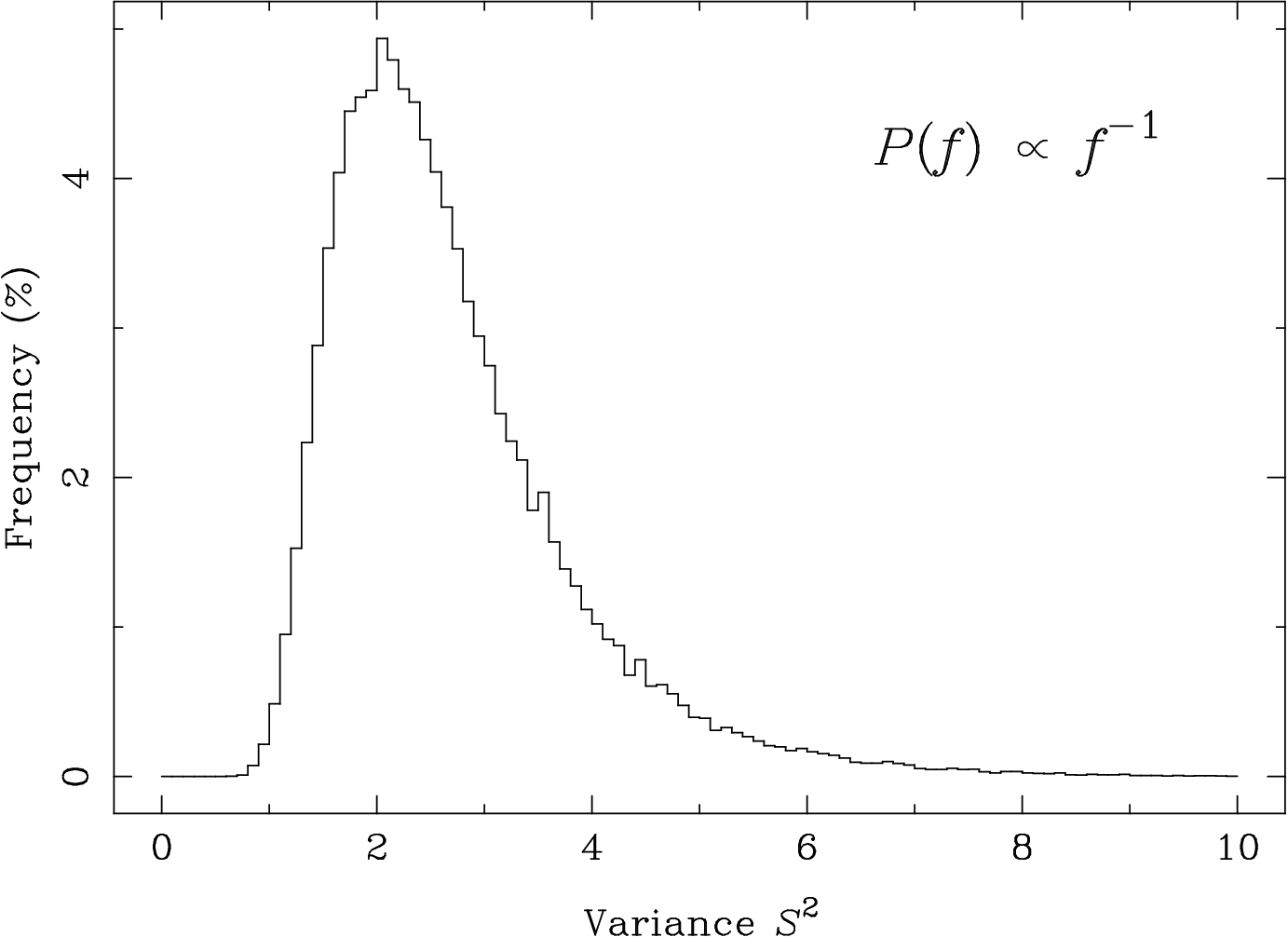}\\
\vspace{0.2 cm}
   \includegraphics[width=5.0 cm, angle=270]{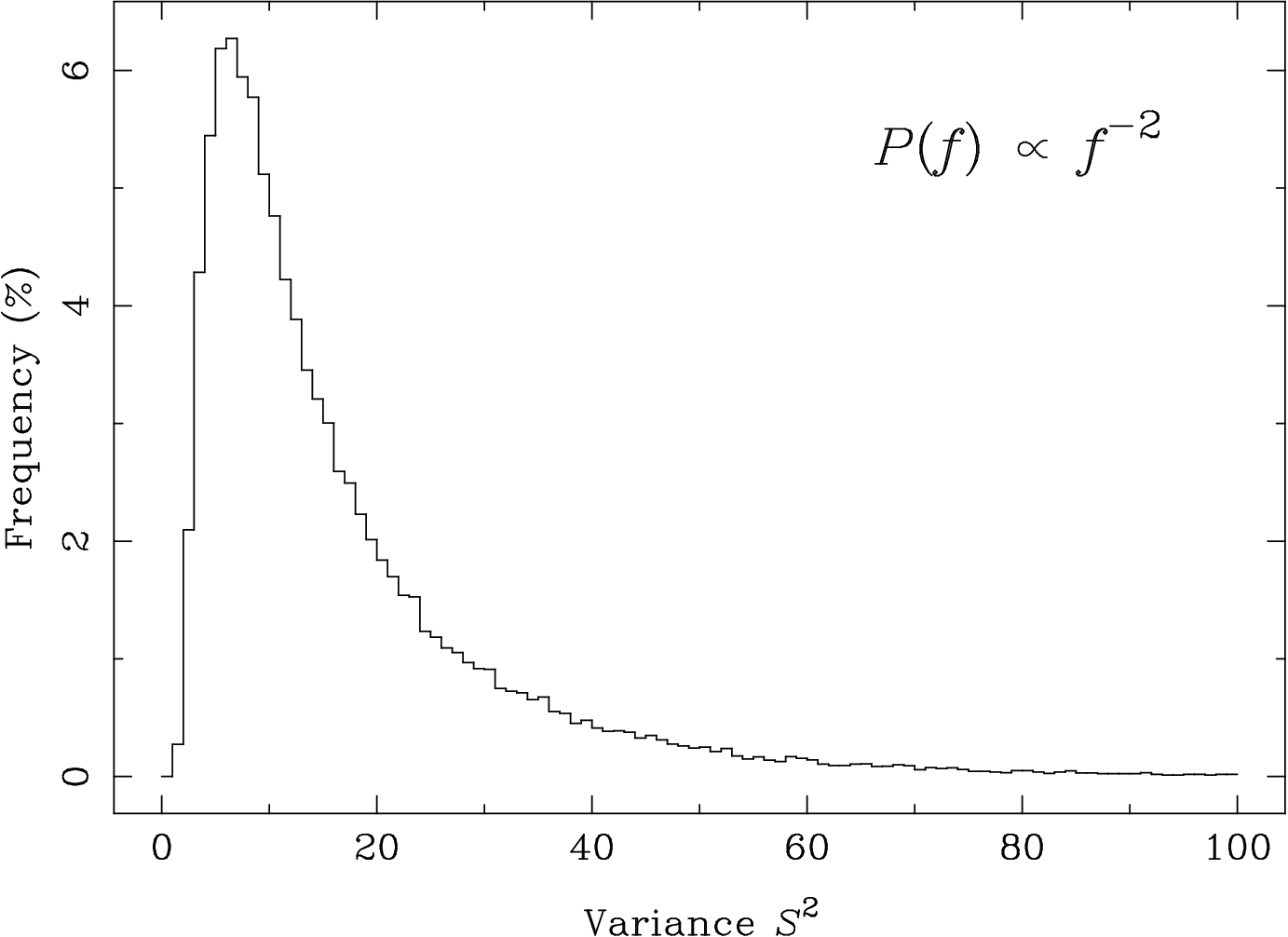}
  \caption{Distribution of variances in time series with three
    different PSD shapes: $f^{0}$ (top), $f^{-1}$ (middle)
and
$f^{-2}$ (bottom). Each distribution is derived from 50,000
realisations. As the PSD gets
steeper the distribution of variances becomes less Gaussian and more
like a $\chi^{2}$ distribution with a low effective degrees of
freedom. 
\label{fig:fvar_dist1}
}
\end{center}
\end{figure}

Fluctuations in the statistical moments (such as mean and variance) of
a light curve are intrinsic to red noise processes. 
Therefore, even in the absence of measurement errors (e.g. no Poisson
noise) the means and variances of two light curves produced
by exactly the same process can be significantly different.
This can be seen in  Fig.~\ref{fig:curve1} (panels 2 and 3), 
where each 20 point segment of the light curve shows a different
mean and variance.
These random fluctuations in variance are however governed by the
normal statistical rules of noise processes and can thus be understood in 
a statistical sense.

Any given series is only one realisation of the processes and
its periodogram will show the scatter predicted by
equation~\ref{eqn:pds_scatter}.  
The integrated periodogram (which gives the variance;
equation~\ref{eqn:correspondence}) will therefore be randomly 
scattered around the true value for the PSD of the process.
The variance in a specific time series is given by
\begin{equation}
\label{eqn:sum_var}
S^{2} = \frac{1}{N \Delta T} \sum_{i=1}^{N/2} \mathcal{P}(f_{i})
\chi_{2}^{2}/2, 
\end{equation}
i.e. the variance of a given realisation  is a sum of $\chi_{2}^{2}$
distributions weighted by the PSD\footnote{As noted earlier, the
periodogram at the Nyquist frequency is actually distributed as
$\chi_{1}^{2}$ for even $N$.  However, for large $N$ this will make a
negligible difference to the sum (cf. equation 2.9 of van der Klis,
1989).}.  (This assumes biases such as red noise leak are not
significant. If this is not true then these biases will further
distort the distribution of variances.)  
It is the expectation value of the variance that is representative of the
integrated power in the PSD, and thus the average amplitude of the
variability process (eqn.~\ref{eqn:correspondence1}).
Thus, while the expectation value of the variance is equal to the
integrated PSD, each realisation (time series) of the same process will show a
different variance even if the parent variability process is stationary.
This is particularly important for steep PSD time series (i.e. weakly
non-stationary data) since the variance does not converge as 
more data are collected. Only if the time series spans timescales 
on which the integrated power converges at low frequencies
(i.e. $\alpha <1$) will the variance converge as the length of
time series increases. 

These points are illustrated by Fig.~\ref{fig:fvar_dist1}, which shows
the distributions of variances in random time series with three
different PSD slopes ($\alpha=0,1,2$).  This plot was produced by
generating 50,000 random time series (each 100 points long) for each
PSD slope and measuring the variance of each one.  The scatter in the
variance is entirely due to random fluctuations between different
realisations because the PSD normalisation was kept fixed and no
instrumental noise was added.  The shape of the
distribution of variances can be seen to depend on the PSD slope.

Consider a white noise process ($\alpha=0$, i.e.
$\mathcal{P}(f)=const$). The periodogram of each realisation is
randomly scattered 
around its flat PSD.  Following equation~\ref{eqn:sum_var} the
variance is  simply the sum of the $N/2$ $\chi_{2}^{2}$-distributed
powers in the periodogram, and these are evenly weighted (PSD is
constant).  The sum of $N/2$ $\chi_{2}^{2}$ distributions follows a
$\chi_{N}^{2}$ distribution. This tends to a normal distribution as $N$
increases. Thus the variance of a white noise process is
approximately normally distributed (as can be seen in
Fig.~\ref{fig:fvar_dist1}) and converges as $N$ increases.   The
fractional standard deviation of $\chi_{N}^{2}$ is given by
$\sqrt{2/N}$, so for the 100 point light curves used the (1$\sigma$)
fractional width of the variance distribution is $\approx 14.1$ per cent,
in agreement with the simulations (Fig.~\ref{fig:fvar_dist1}, top panel).

For time series with steeper PSDs the lower frequency periodogram
points contribute more strongly to the sum than the higher frequency
points. The variance of such a time series is therefore
dominated by a few low frequency powers\footnote{The windowing effects
  mentioned above mean that 
fluctuations at powers above and below the frequency range of the periodogram
may also affect the variance.} and thus resembles a
$\chi_{\nu}^{2}$ distribution with low `effective degrees of
freedom.'  The distribution of variances in red
noise data is dependent on the underlying PSD and is, in general,
non-Gaussian (Fig.~\ref{fig:fvar_dist1}). The fractional
standard deviation of $\chi_{\nu}^{2}$ is $\sqrt{2/\nu}$, which tends
to unity as the PSD gets steeper (i.e. as effective $\nu \rightarrow 2$).
Thus the largest fluctuations in variance (up to a limit of $\sim 100$
per cent rms) are expected to result from very steep PSD slopes.

\section{Intrinsic scatter in variance}
\label{sect:errors_intrinsic}

As discussed above, when examining AGN light curves one should expect
random changes in the mean and variance with time (between segments of
a long observation or between observations taken at different epochs).
This is true even if the measurement errors are zero and does not
depend on the number of data points used (due to the weak
non-stationarity).  However, it is also  possible that the underlying
process responsible for the variability itself changes with time
(e.g. the PSD changes), in which case the variability is
non-stationary in a more meaningful sense -- `strongly
non-stationary.' Such changes in the variability process could
provide  insight into the changing physical conditions in the nuclear
regions. On the other hand the random changes expected for a red noise
process yield no such physical insight. The question thus arises: how
does one tell, from a set of time series of the same source, whether
they were produced by a strongly non-stationary process? In other
words, is it possible to differentiate between differences in variance
caused by real changes in the variability process (physical changes in
the system), and random fluctuations expected from red noise (random
nature of the process)?

If the process responsible for the variability observed in a given
source is stationary then its PSD is constant in time. The expectation value
of the absolute (un-normalised) variance will therefore be the
same from epoch to epoch, but the individual variance estimates will
fluctuate as discussed in section~\ref{sect:stochasticity}.  This
makes it difficult to judge, from just the variances of two light
curves taken at two epochs, whether they were produced by a stationary
process. Given sufficient data it is, however,  possible to test
whether the expectation values of the variance (estimated from an
ensemble of light curves) at two epochs are consistent with a
stationary process.

\subsection{Comparing PSDs}

The methods most frequently employed involve comparing the PSDs
(estimated from the binned periodogram) at different epochs.  If the
PSDs show significant differences (at a given confidence level) 
the variability process can be said to be strongly
non-stationary. As an example of this, the PSDs of X-ray binaries
evolve with time, and the way in which the variability properties
evolve provides a great deal of information on the detailed workings
of these systems (see e.g. Belloni \& Hasinger 1990; Uttley \& M$^{\rm
c}$Hardy 2001; Belloni, Psaltis \& van der Klis 2002; Pottschmidt \et
2002).

Papadakis \& Lawrence (1995) suggested a method suitable for testing
whether large AGN datasets display evidence for strongly
non-stationary variability. Again this method works by comparing the
PSDs from different time intervals, in this case by determining
whether the differences between two periodograms are consistent with
the scatter expected based on equation~\ref{eqn:pds_scatter}. In
particular, they define a statistic $s$ based on the ratio of two
normalised periodograms. If $s$ deviates significantly from its
expected value for stationary data (if $\langle s \rangle =0 $) 
then the
hypothesis that the data are stationary can be rejected (at some
confidence level).

\subsection{Comparing variances}

A different approach is compare variances $S_i$ derived from $M$
observations of the same source (either segments of  one long
observation or separate short observations).  In order to test whether
the $S_i$ differ significantly (i.e. more than expected for a red
noise process) a measure of the expected scatter is required. This
error could be obtained directly from the data (by measuring the
standard deviation of multiple estimates) or through simulations
(based on an assumed PSD shape)\footnote{It is assumed that the data segments being compared
have identical sampling (same bin size and observation length). 
Every effort should be made to ensure this is the case,
e.g. by clipping the segments to the same length.
The variances will then be calculated over the same
range of timescale (frequencies).
As the variance can increase rapidly with timescale in 
red noise data this is most important for steep PSD data such as
AGN light curves.}.

\subsubsection{Empirical error on variance}
\label{sect:bin_var}

An empirical estimate of the mean and standard deviation of the
variance can be made given $M$ non-overlapping data segments.  The $M$
segments each yield an estimate of the variance\footnote{Ideally each
segment  should contain at least $N \gs 20$ data points in order to
yield a meaningful variance.}, $S_i$. Each of these is an independent
variable of (in general) unknown but identical distribution (unless
the process is strongly non-stationary). The central limit theorem
dictates that the sum of these will become normally distributed as $M$
increases. Therefore by averaging the $M$ variance estimates it is
possible to produce an averaged variance ($\langle S^{2} \rangle$) and
assign an error bar in  the usual fashion (e.g. equation 4.14 of
Bevington \& Robinson 1992).  This gives a robust estimate of the
variance and  the standard deviation of the $M$ variances around the
mean gives an estimate of the uncertainty on the mean variance.

If several sets of data segments are acquired it is therefore
possible to compare the mean variance of each set statistically (since
each has an associated uncertainty).   For example, with two long
\xmm\ observations of the same source, taken a year apart, one could
measure the variance for each observation (by breaking each into short
segments and taking the mean  variance of the segments).  Thus it
would be possible to test whether the variability was stationary
between the two observations.
This method of estimating the mean and standard deviation of
the variance requires a large amount of data. Of order
$N \times M = 20 \times 20 = 400$ data points are needed to produce a single 
well-determined estimate of the mean variance and its error.
A typical \xmm\ observation of a bright Seyfert 1 galaxy ($\sim 40$~ks
duration) is only likely to yield enough data for one estimate of the
mean variance. Thus this method is suitable for testing whether the
mean variance has changed from observation to observation.

\begin{figure}
 \begin{center}
  \rotatebox{-90}{
\includegraphics[width=6.0 cm]{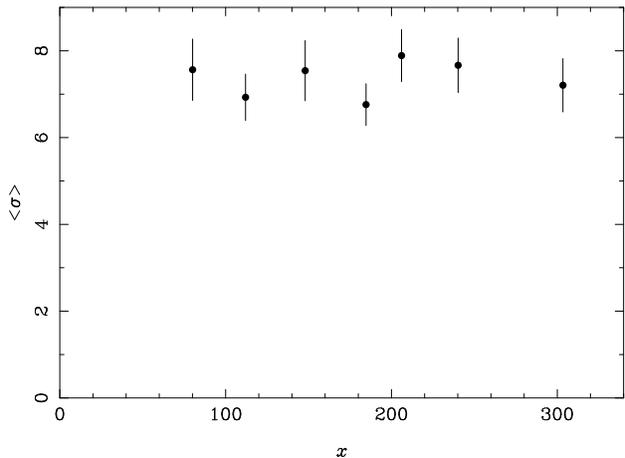}
  }
\caption{
The average rms amplitude ($\sigma=\sqrt{S^{2}}$) as a function of
flux for the simulated 
light curve shown in Fig.~\ref{fig:curve1}.
The individual rms estimates 
were sorted by flux and binned to $M=20$ estimates per bin.
Errors correspond to the error on the mean value.
The amplitude is constant with flux.
\label{fig:curve2}
}
\end{center}
\end{figure}

Fig.~\ref{fig:curve1} (panel 4)
demonstrates this empirically derived mean variance and its error bar
on a long, simulated time series.
These data were produced by calculating the variances $S_i$ in bins of
$N=20$ data points (panel 3) and then averaging $M=20$ variances to
produce a mean variance with error bar (panel 4). These averaged
variances are consistent with constant, as expected; fitting these
data with a constant gave $\chi_{\nu}^{2}=0.84$.
Figure~\ref{fig:curve2} shows the rms amplitude is constant with flux.
These tests indicate that the integrated PSD is consistent with being
constant with time; the
variance does not change significantly from epoch to epoch (or as
function of flux), as expected for a stationary process.

\subsubsection{Estimating the error on the variance through simulations}
\label{sect:sim_table}

The advantage of the above method is that it requires no assumption
about the shape of the PSD, The drawback is that it requires a
substantial amount of data to produce a single, robust variance
estimate. An alternative approach is to estimate the standard
deviation of the variances $S_i$ based on simulations.

Given an assumed shape for the PSD it is possible to  calculate the
distribution of variances expected for a stationary process (see
section~\ref{sect:stochasticity}).  Some example distributions are
shown in Fig.~\ref{fig:fvar_dist1}, which clearly demonstrates how the
distribution depends on the slope of the PSD. The distribution becomes
more normal at flatter slopes and more asymmetric at steep slopes. For
a given PSD shape these distributions are well defined (by
eqn.~\ref{eqn:sum_var}) and can be computed through Monte Carlo
simulations. This makes it possible to estimate limits within
which one would expect the variance to be distributed if the
process is stationary. 

The two primary factors that affect the distribution of variance are
the PSD shape and the sampling of the light curve (the length of the
data segments in the case of contiguously binned light
curves). Table~\ref{table} gives the expected confidence limits  for
four different PSD shapes and five different lengths for the data
segments. These values were computed by simulating one very long light
curve with the assumed PSD shape and breaking it into $1000$ separate
segments (of specified length). The variance within each segment was
measured and the distribution of the $1000$
variances was calculated. The $90$ per cent confidence interval
was calculated by finding the $5$th and $95$th percentiles of the
variance distribution (in general these upper and lower bounds will
differ because the distribution is asymmetric).  The numbers given in
the table are the boundaries of the $90$ and $99$ per cent confidence
regions estimated by averaging the results from $50$ runs. The limits
are given in terms of $\pm \Delta \log(S^2)$ because they are multiplicative.
That is, from a particular realisation the variance is expected to be 
scattered within some factor of the true variance (for which the
absolute normalisation is irrelevant). The factors are tabulated in
terms of their logarithms (since multiplicative factors in
linear-space become additive offsets in log-space).

The PSD used for the simulations was chosen to match that expected for
AGN, i.e. a steep power-law at high frequencies (with a slope of
$\alpha = 1.0,1.5,2.0,2.5$) breaking to a flatter slope ($\alpha = 1.0$) at
low frequencies. The frequency of the break was fixed to be $10^{-3}$,
in other words the break timescale was $1000$ times the bin size.
(The absolute size of the time bins is arbitrary in
the simulations. When comparing the simulated results to real data
sampled with e.g. $25$~s time resolution, the break timescale in the 
simulated PSD is thus $25$~ks.)

\begin{table}
\caption{
Expected scatter in variance estimates. 
The $90$ and $99$ per cent intervals are presented in terms of 
$\pm \Delta \log(S^2)$. (The $99$ per cent interval is given in {\bf bold}.)
The boundaries were calculated from Monte Carlo simulations
of light curves. The PSD was chosen to be a broken power-law 
with a slope of $\alpha = 1$ below the break (at a frequency
$10^{-3}$) and a slope above the break of $\alpha = 1.0,1.5,2.0,2.5$. 
The simulated data segments were chosen to be 
$10, 20, 50, 100, 1000$ points long (the timescale of the break
in the PSD being at $1000$, in arbitrary units).  }
\centering
\label{table}            
\begin{tabular}{@{}l|rrrrr@{}}
\hline
          & \multicolumn{5}{c}{Number of data points} \\
$\alpha$  &  10         &  20         &  50         & 100         &  1000       \\
\hline
1.0       & {\bf -0.84}       &  {\bf -0.58}      &  {\bf -0.40}      &  {\bf -0.32}      &  {\bf -0.19}      \\
          & -0.50       &  -0.36      &  -0.26      &  -0.22      &  -0.13      \\
          & +0.33       &  +0.28      &  +0.23      &  +0.20      &  +0.15      \\
          & {\bf +0.53}       &  {\bf +0.46}      &  {\bf +0.39}      &  {\bf +0.35}      &  {\bf +0.27}      \\
\hline
1.5       & {\bf -0.96}       &  {\bf -0.75}      &  {\bf -0.62}      &  {\bf -0.57}      &  {\bf -0.45}      \\
          & -0.61       &  -0.50      &  -0.43      &  -0.40      &  -0.32      \\
          & +0.39       &  +0.36      &  +0.34      &  +0.33      &  +0.28      \\
          & {\bf +0.65}       &  {\bf +0.61}      &  {\bf +0.58}      &  {\bf +0.57}      &  {\bf +0.49}      \\
\hline
2.0       & {\bf -1.16}       &  {\bf -1.01}      &  {\bf -0.93}      &  {\bf -0.90}      &  {\bf -0.72}      \\
          & -0.78       &  -0.71      &  -0.67      &  -0.66      &  -0.50      \\
          & +0.46       &  +0.45      &  +0.44      &  +0.43      &  +0.36      \\
          & {\bf +0.75}       &  {\bf +0.73}      &  {\bf +0.72}      &  {\bf +0.72}      &  {\bf +0.59}      \\
\hline
2.5       & {\bf -1.49}       &  {\bf -1.37}      &  -{\bf 1.31}      &  {\bf -1.28}      &  {\bf -0.92}      \\
          & -1.03       &  -0.98      &  -0.95      &  -0.92      &  -0.63      \\
          & +0.52       &  +0.52      &  +0.51      &  +0.50      &  +0.40      \\
          & {\bf +0.83}       &  {\bf +0.83}      &  {\bf +0.82}      &  {\bf +0.80}      &  {\bf +0.66}      \\
\hline
\end{tabular}
\end{table}

\begin{figure}
 \begin{center}
  \rotatebox{-90}{
\includegraphics[width=6.0 cm]{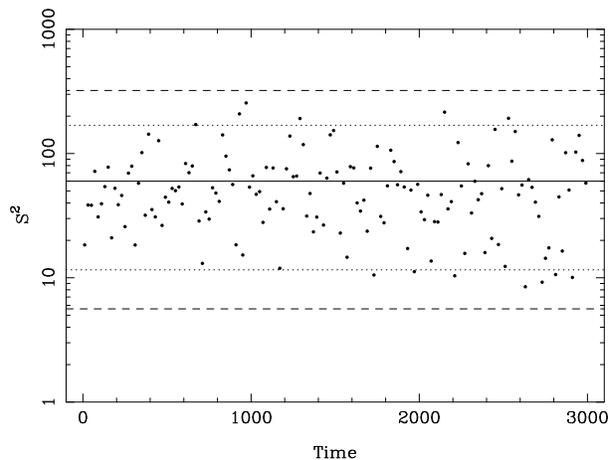}
  }
\caption{
Variance of the simulated data shown in Fig.~\ref{fig:curve1} (panel
3) with the $90$ (dotted line) and $99$ per cent (dashed line)
confidence intervals marked (as calculated in
section~\ref{sect:sim_table}). 
Clearly the variances fall within these
limits, as expected for a stationary process.
The solid line marks the mean variance.
}
\label{fig:sim_result}
\end{center}
\end{figure}

The numbers given in the table provide an approximate prescription for
the expected scatter in the variance of a stationary process with a
red noise PSD similar to that of AGN.  The simulated light curve
shown in Fig.~\ref{fig:curve1} was used to demonstrate the use of this
table. In this case the PSD is know to have a slope $\alpha = 2$, and 
the variances (shown in panel 3 of Fig.~\ref{fig:curve1}) were
calculated every $20$ points. Therefore, the $90$ interval for
the expected variance is given by $\log(S^2)_{-0.71}^{+0.45}$. 
Taking the mean variance as the expectation value for $S^2$, this translates
to $S^2 = 59.9 ~(11.7 - 168.8)$. The interval boundaries were calculated
by converting the logarithmic value into a linear factor and multiplying by
the sample mean (assumed to represent the true variance). This
interval is shown on Fig.~\ref{fig:sim_result} by the dotted 
lines. The corresponding $99$ per cent confidence interval is also marked.

As expected the individual variances fall within the expected
region. However, the $90$ per cent region spans an order of magnitude 
in variance. Thus even order of magnitude differences in
variance between short sections of a light curve are to be expected
and do not necessarily indicate that the underlying process is not
stationary. Subtle changes in the PSD will thus be difficult to
detect by examining the raw variances as the intrinsic scatter is
so large. Such changes could be revealed by comparing averaged
variances or comparing the PSDs as described above.

\section{Effect of measurement errors}
\label{sect:errors_poisson}

\subsection{Excess variance and $F_{\rm var}$}

The datasets considered thus far have been ideal, in the sense
that they are free from flux uncertainties. In real life,
however, a light curve $x_i$ will have finite uncertainties
$\sigma_{{\rm err},i}$ due to measurement errors (such as Poisson
noise in the case of an X-ray photon counting signal).
These uncertainties on the individual flux measurements
will contribute an additional variance. This leads to
the use of the `excess variance' (Nandra \et 1997; Edelson \et 2002)
as an estimator of the {\it intrinsic} source variance. 
This is the variance after
subtracting the contribution expected from measurement errors
\begin{equation}
\label{eqn:excess_variance}
\sigma_{\rm{XS}}^{2} = S^{2} - \overline{\sigma_{\rm{err}}^2},
\end{equation}
where $\overline{\sigma_{\rm{err}}^{2}}$ is the mean square error
\begin{equation}
\label{eqn:mean_error}
\overline{\sigma_{\rm{err}}^{2}} = \frac{1}{N}\sum_{i=1}^{N} \sigma_{{\rm
err}, i}^{2}.
\end{equation}
The normalised excess variance is given by
$\sigma_{\rm{NXS}}^{2}=\sigma_{\rm{XS}}^{2}/\bar{x}^{2}$ and
the fractional root mean square (rms) variability amplitude
($F_{\rm{var}}$; Edelson, Pike \& Krolik 1990; Rodriguez-Pascual \et
1997) is the square root of this, i.e.
\begin{equation}
\label{eqn:fvar}
F_{\rm{var}} = \sqrt{ \frac{S^{2} -
\overline{\sigma_{\rm{err}}^{2}}}{\bar{x}^{2}}}.
\end{equation}

The statistic $F_{\rm var}$ is often chosen in preference to
$\sigma_{\rm NXS}^{2}$ although the two convey exactly the
same information. $F_{\rm var}$ is a linear statistics and
can therefore give the rms variability amplitude in per centage
terms. The choice of whether to quote $F_{\rm var}$ or $\sigma_{\rm
  NXS}^{2}$ is usually purely one of presentation. It is worth
noting that the Monte Carlo results given in
section~\ref{sect:sim_table}, to estimate the expected scatter on
the variance, can also be applied to its square root. The expected
boundaries of the confidence region of
the logarithm of the rms is approximately half those of the
variance. Specifically, 
$\Delta \log(\sigma) \approx \Delta
\log(S^{2}) /2$ and similarly 
$\Delta \log(F_{\rm var}) \approx \Delta
\log(\sigma_{\rm NXS}^{2}) /2$.

\subsection{Spectral variability}
\label{sect:spec_var}

An X-ray light curve of an AGN can be split into different energy
bands. The light curves in each band will be strictly simultaneous and
can be used to test whether the X-ray variability is  a function of
energy. For example, one might examine the ratio of a soft band light
curve to a hard band light curve. The statistical significance of any
variations in the ratio can be quantified by propagating the
measurement errors and applying an appropriate test, such as the
$\chi^{2}$ test of the constant ratio hypothesis\footnote{This does,
of course, assume the light curves have been binned sufficiently for
the error bars to be approximately Gaussian.}. If the ratio shows
variations greater than those expected from the errors then the two
light curves are intrinsically different and the source does indeed
show spectral variability. Such changes  in the energy spectrum with
time can in principle provide valuable clues to the nature of the
X-ray source. This test does not provide any quantitative description
of the spectral variability.

Another tool for investigating spectral variability is the rms
spectrum, i.e. the rms variability amplitude (or $F_{\rm var}$) as a
function of energy. See e.g. Inoue \& Matsumoto (2001), Edelson \et
(2002), Fabian \et (2002) and Schurch \& Warwick (2002) for some
examples of rms spectra from AGN.  However, when examining rms spectra
it is often not clear whether changes in the amplitude with energy
reflect real energy-dependence of the intrinsic variability amplitude
or are caused by random errors in the light curves. The finite
measurement errors on the individual fluxes (e.g. due to Poisson
noise)  will introduce some uncertainty in the estimated rms
amplitudes.  An estimate of this uncertainty would help answer the
question posed above, namely whether features in rms spectra are the
result of random errors in the data or represent spectral variations
intrinsic to the source.

The problem of how to assess the uncertainty on the excess variance
(or $F_{\rm var}$) is a long-standing one (e.g. Nandra \et 1997;
Turner \et 1999; Edelson \et 2002). The standard error formulae
presented in the literature (e.g. Turner \et 1999; Edelson \et 2002) 
are formally valid in the case of un-correlated Gaussian processes. 
Typically AGN light curves at different X-ray energies are strongly 
correlated and are not Gaussian. However, when searching for subtle
differences in amplitude between simultaneous and correlated light
curves it may be more useful to have an indication of the uncertainty
resulting from the finite flux errors.

\subsubsection{Uncertainty on excess variance due to measurement errors}

A Monte Carlo approach was used to develop a prescription of the
effect of measurement errors on estimates of $F_{\rm var}$ (and
$\sigma_{\rm{NXS}}^2$). A short red noise light curve was generated.
Poisson noise was added (i.e. the individual flux measurements were
randomised following the Poisson distribution) and the excess variance
was recorded. The fluxes of the original light curve were randomised
again and the excess variance recorded, this was repeated many times.
The distribution of excess variances was then used to determine the
uncertainty in the variance estimate caused by Poisson noise. Full
details of the procedure are given in Appendix~\ref{sect:MC}

For these simulations it was found that the error on
$\sigma_{\rm{NXS}}^2$ decreases as the S/N in the light curve is
increased according to:
\begin{equation}
err(\sigma_{\rm{NXS}}^{2}) = \sqrt{
\left\{
\sqrt{\frac{2}{N}}\cdot\frac{\overline{\sigma_{\rm{err}}^{2}}}{\bar{x}^{2}}
\right\}^{2}
+
\left\{ \sqrt{\frac{\overline{\sigma_{\rm{err}}^{2}}}{N}}\cdot \frac{ 2
F_{\rm{var}}}{\bar{x}}  \right\}^{2} .
}
\label{eqn:xs_error}
\end{equation}
See appendix~B for details of this equation and its equivalent in
terms of $F_{\rm{var}}$.

As this only accounts for the effect of flux measurement errors (such
as Poisson noise) in a given light curve
it can be used to test whether two simultaneously observed light
curves of the same source, but in different bands, show consistent
amplitudes. A demonstration of this using real data is given in the
following section. This uncertainty does not account for the
random scatter intrinsic to the red noise process, therefore the
absolute value of the rms spectrum will change between realisations
(i.e. from epoch to epoch). But if a source shows achromatic variability
then the values of $F_{\rm{var}}$ calculated in each energy band (at
a given epoch) should match to within the limits set by the Poisson
noise (i.e. the fractional rms spectrum should be constant to within
the uncertainties given by the above equation). 
Differences in $F_{\rm{var}}$ significantly larger than these would
indicate that the source variability amplitude is a function of
energy. This would then mean the PSD amplitude/shape
is different in different energy bands, or there are multiple
spectral components that vary independently.

The above uncertainty estimates can be used to test the hypothesis
that the source variability is achromatic. If significant differences
between energy bands are detected (as in the case of \mrk\ presented
below) then these errors should not be used to fit the rms spectrum. 
The assumption that the differences are due only to measurement errors
is no longer the case. In such situations the light curves in adjacent
energy bands are likely to be partially correlated and so 
$\chi^{2}$-fitting of the rms spectrum is not appropriate. The
differences in excess variance will be a combination of intrinsic
differences and measurement errors. Their uncertainty 
will therefore be more difficult to quantify.

\section{Case study: An \xmm\ observation of \mrk}
\label{sect:mrk766}

\begin{figure}
\begin{center}
   \includegraphics[width=8.5 cm, angle=0]{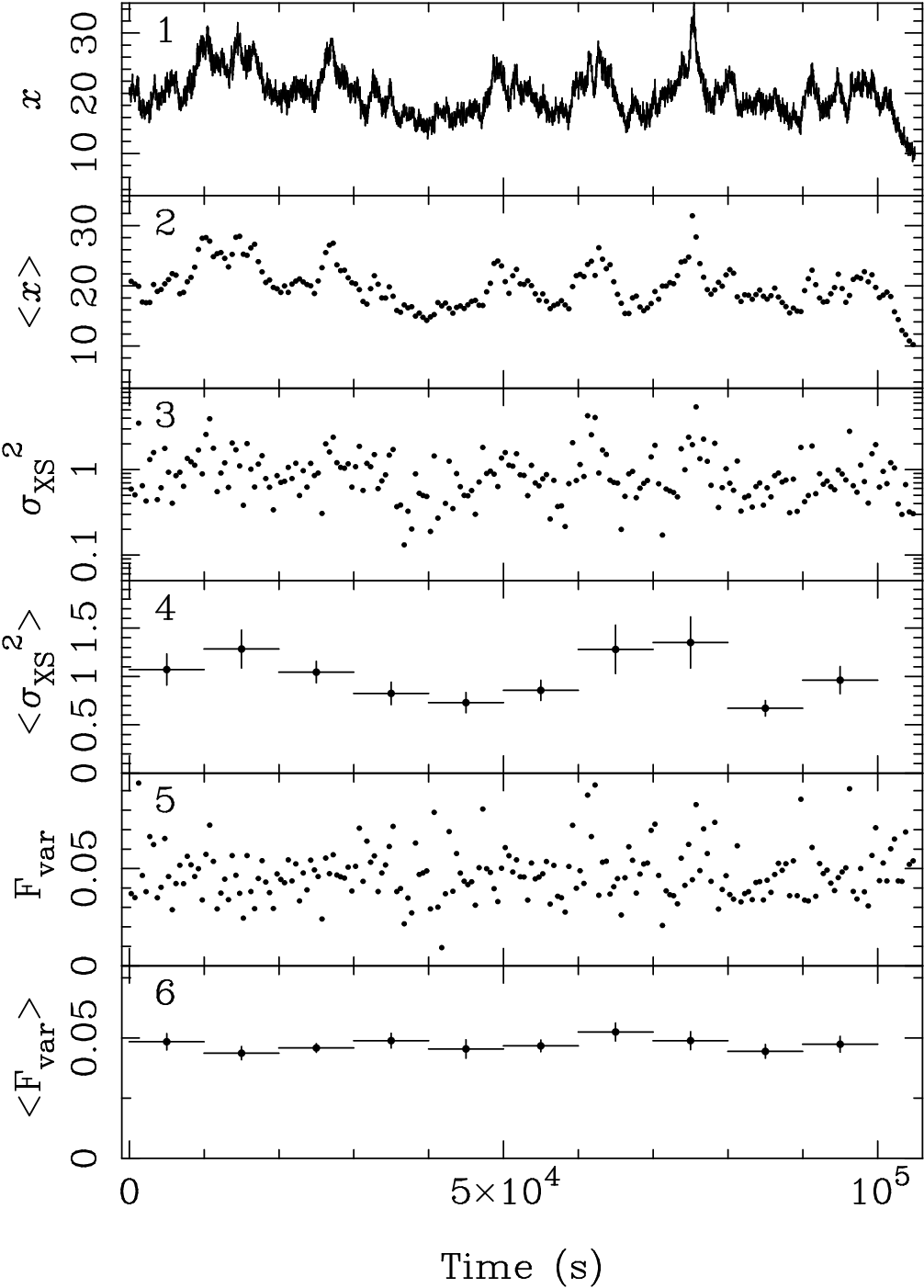}
\end{center}
\caption{
Top panel: 0.2--10.0 keV pn light curve of \mrk\ (with 25~s bins, in units of ct s$^{-1}$).
Panel 2 and 3: mean count rate and excess variance measured from segments of 20 points. 
Panel 4: averaged normalised excess variance measured by binning
the individual variance estimates into groups of 20.
This average variance is inconsistent with constant.
Panel 5: fractional rms amplitude measured from segments of 20 points.
Panel 6: averaged fractional rms amplitude measured by binning
the individual amplitudes into groups of 20.
This average fractional amplitude is consistent with constant.
This contrasts with the situation shown in Fig.~\ref{fig:curve1}
\label{fig:mrk766_1}
}
\end{figure}

In this section a long ($\sim 10^5$~s) \xmm\ observation of the
bright, variable Seyfert 1 galaxy Markarian 766 is used to  illustrate
the points discussed above.  The data were obtained from the \xmm\
Data  Archive\footnote{{\tt
http://xmm.vilspa.esa.es}}.
Details of the observation are discussed in Mason \et (2003) and an
analysis of the PSD is presented in Vaughan \& Fabian (2003).

\subsection{Observation details}

\mrk\ was observed by \xmm\ (Jansen \et 2001) over the period 2001 May
20 -- 2001 May 21 (rev. 265). The present analysis is restricted to
the data from the pn European Photon Imaging Camera (EPIC),
which was operated in small window mode.
Extraction of science products from the Observation
Data Files (ODFs) followed standard procedures using the \xmm\ Science
Analysis System (SAS) v5.3.3.  Source data were extracted from a
circular region of radius 35 arcsec from the processed image and
only events corresponding to patterns 0--4 (single and double
pixel events) were used. Background events
were extracted from regions in the small window least effected by
source photons, these showed that the background rate increased
dramatically during the final $\sim 1.5\times 10^3$~s of the observation. This
section of the data was excluded, leaving $1.05\times 10^5$~s of uninterrupted
data.  The light curves were corrected for telemetry drop outs
(less than 1 per cent of the total time) and background
subtracted. The errors on the light curves were calculated by
propagating the Poisson noise.

\subsection{Stationarity of the data}
\label{sect:mrk766_stationary}

\begin{figure}
 \begin{center}
  \rotatebox{-90}{
\includegraphics[width=6.0 cm]{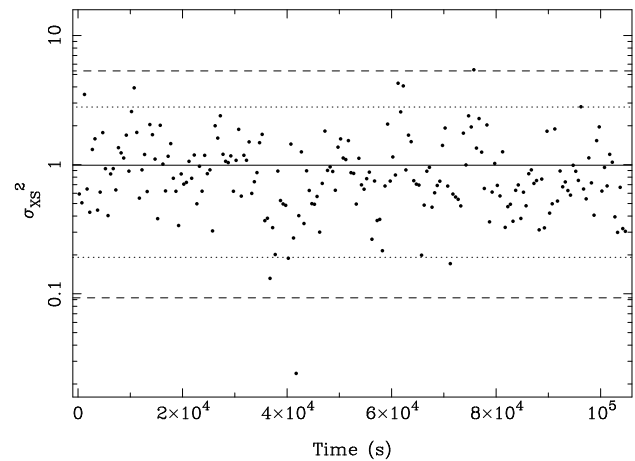}
  }
\caption{
Excess variance of the \mrk\ data shown in Fig.~\ref{fig:mrk766_1} (panel
3) with the $90$ (dotted line) and $99$ per cent (dashed line)
confidence intervals marked (as calculated in
section~\ref{sect:sim_table}). 
The variances fall within these
limits, as expected for a stationary process.
The solid line marks the mean variance.
}
\label{fig:766_result}
\end{center}
\end{figure}

\begin{figure}
 \begin{center}
  \rotatebox{-90}{
\includegraphics[width=6.0 cm]{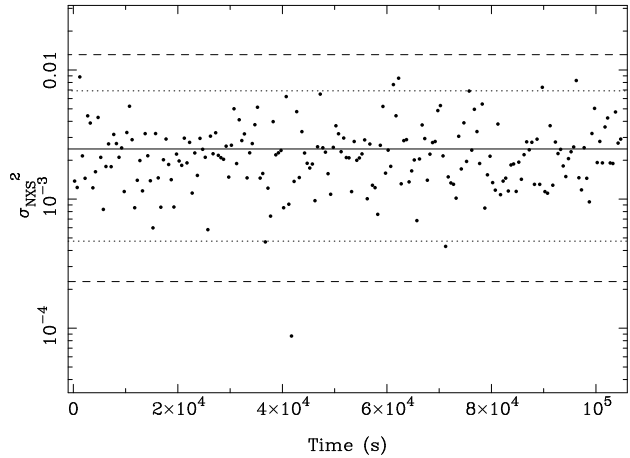}
  }
\caption{
As for Fig.~\ref{fig:766_result} but using the normalised excess
variance of the \mrk\ data.
}
\label{fig:766_result2}
\end{center}
\end{figure}

The broad band (0.2--10~keV) light curve extracted from the pn  is
shown in Fig.~\ref{fig:mrk766_1} (panel 1). As was the case for the simulated
data shown in  Fig~\ref{fig:curve1}, the mean and variance (calculated
every 20 data points)\footnote{These correspond to `instantaneous'
estimates of the source variance on timescales of $50 - 500$~s.}  show
changes during the length of the observation (panels 2 and 3).  The
expected range for the excess variance, calculated using the results of
section~\ref{sect:sim_table} (and assuming a PSD slope of $\alpha =
2.0$), is marked in figure~\ref{fig:766_result}.
Fig.~\ref{fig:766_result2} shows the same data in terms of normalised
excess variances. Neither of these show fluctuations larger than
expected for a stationary process. But given the large expected
scatter this is a rather insensitive test.
In the case of the \mrk\ light curve however, there are sufficient data
to examine variations of the average variance with time, allowing a more
sensitive test for non-stationarity.

By averaging the excess variance estimates (in time bins containing 20
excess variance estimates)  significant changes in the variance with
time are revealed (panel 4).  This contrasts with the simulated data
shown in Fig.~\ref{fig:curve1}.  The binned excess variance is
inconsistent with a constant hypothesis: fitting with a constant gave
$\chi^{2} = 23.1$ for $9$ degrees of freedom ($dof$), rejected at $99$
per cent confidence.  The average variance is therefore changing with
time, indicating the variability is strongly non-stationary.

\begin{figure}
 \begin{center}
   \includegraphics[width=6.0 cm, angle=270]{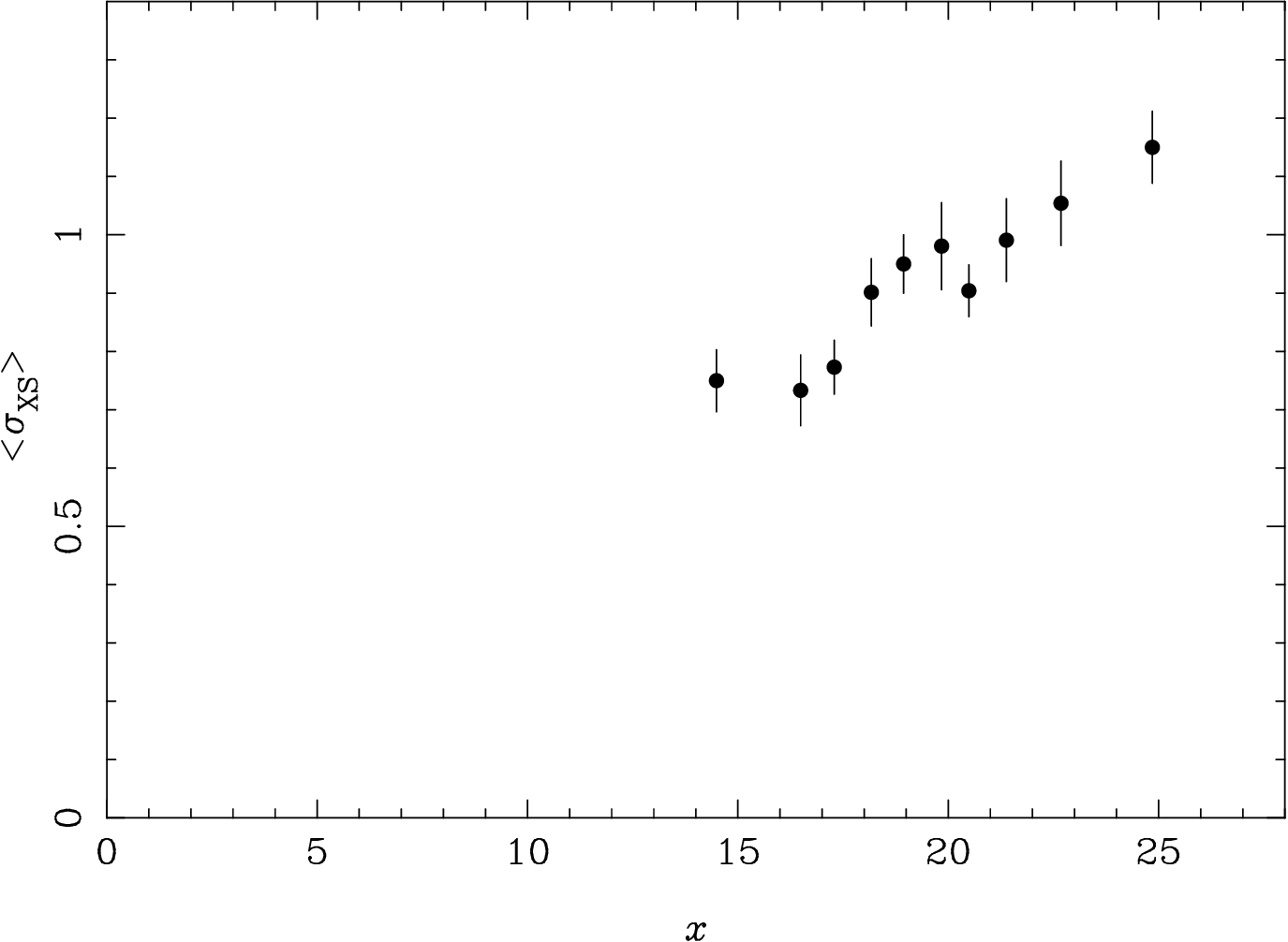}\\
\vspace{0.5 cm}
   \includegraphics[width=6.0 cm, angle=270]{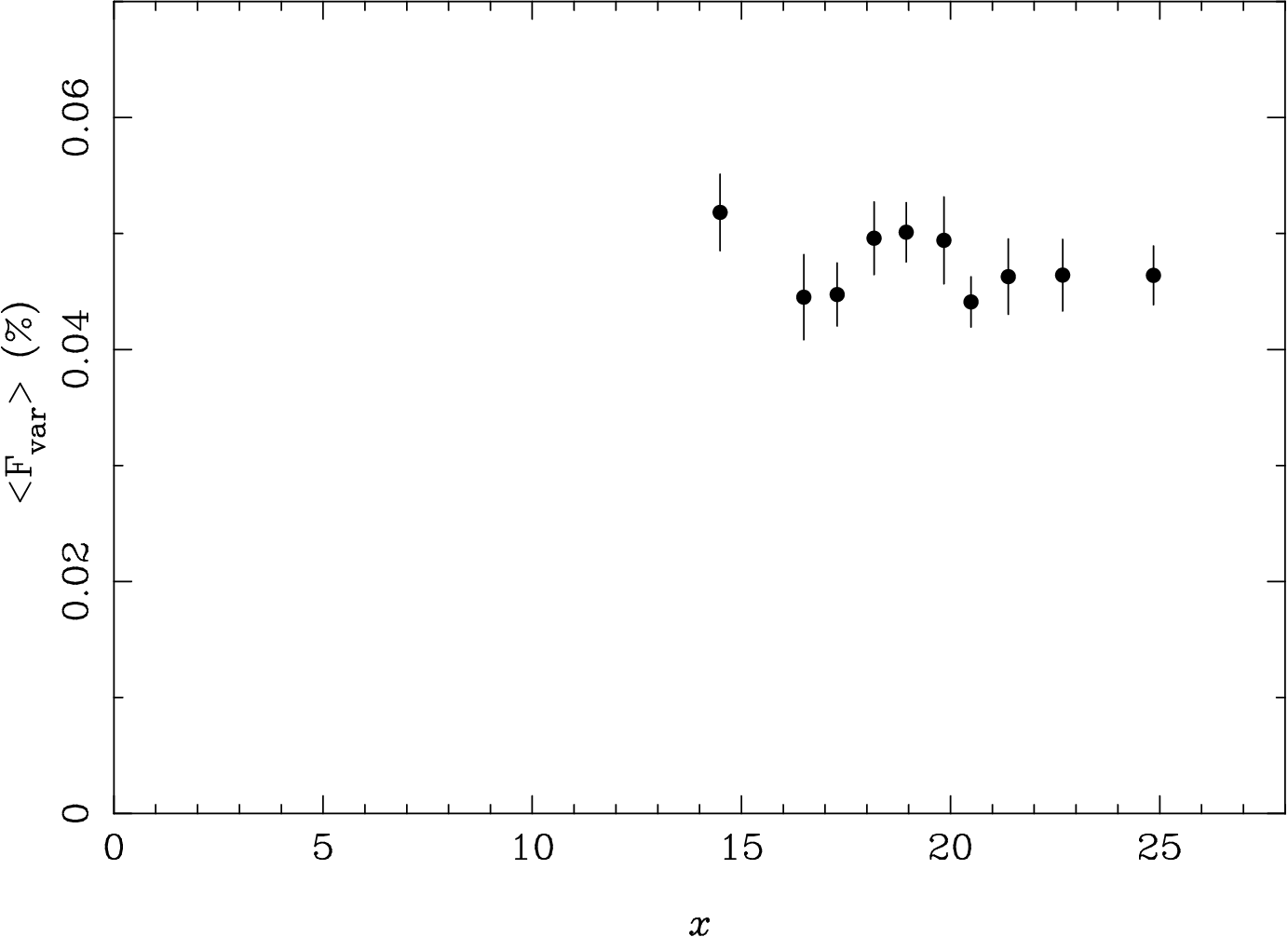}
\caption{
Top panel: The average absolute rms amplitude ($\sqrt{\sigma_{\rm XS}^{2}}$) 
as a function of flux for the \mrk\  light curve (compare
with Fig.~\ref{fig:curve2}).  Bottom panel: The average {\ it
fractional} rms amplitude ($\sqrt{\sigma_{\rm NXS}^{2}}$) as a function
of flux.  Clearly the absolute rms amplitude is a function of flux,
but this dependence is removed in the fractional rms.
\label{fig:mrk766_rms}
}
\end{center}
\end{figure}

A careful inspection of Fig.~\ref{fig:mrk766_1} (panels 3 and 4) shows the individual
variance estimates have 
a tendency to track the source count rate. This is difficult to 
discern from the individual variances (panel 3), due to the larger
intrinsic scatter, but much clearer in the averaged variances (panel 4).
This can be seen clearly in
Fig.~\ref{fig:mrk766_rms} (top panel) where the rms amplitude ($\sqrt{\sigma_{\rm
 XS}^{2}}$) is shown as a function of count rate. To produce
this plot the individual rms estimates (Fig.~\ref{fig:mrk766_1}, panel
3) were sorted by count rate and
binned by flux (such that there were $20$ estimates per bin).
The error on the mean rms was calculated in the standard fashion (see
above). 
This indicates that the source does show a form of genuine
non-stationarity: the absolute rms amplitude of the variations increases,
on average, as the source flux increases. This effect has been noted in other
Seyferts (Uttley \& M$^{\rm c}$Hardy 2001; Edelson \et 2002; Vaughan \et 2003) and is due
to a linear correlation between rms and flux (see Uttley \et in prep.
for further discussion of this effect). Non-stationarity of this
form can be `factored out' by using the normalised amplitude
($F_{\rm var}$ or $\sigma_{\rm NXS}^{2}$) instead of the absolute
values. Normalising each variance (or rms) estimate by its local 
flux removes this trend.
The bottom panel of Fig.~\ref{fig:mrk766_rms} shows 
that $F_{\rm var}$ is indeed constant with flux (fitting a 
constant gave $\chi^{2} = 7.8$ for $9 ~ dof$). 
Fig.~\ref{fig:mrk766_1} (panels 5 and 6) shows $F_{\rm var}$
and its average as a function of time; the average is consistent
with staying constant ($\chi^{2} = 5.8$ for $9 ~ dof$).
The variability of \mrk\ does show genuine (strong) non-stationarity, in the
sense that the absolute rms increases linearly with flux, but this
trend can be removed by using normalised units -- $F_{\rm var}$ (and
therefore the normalised excess variance) is
consistent with being constant (with time and flux).

The above analysis suggests that, after accounting for the effect of the
rms--flux correlation, there is no other evidence for strong non-stationarity
in the rapid variability of \mrk.
This was confirmed using the $s$-test of Papadakis \& Lawrence (1995; see
their Appendix A). A periodogram was calculated for 
three consecutive segments of $3.4\times 10^{4}$~s duration, and normalised to fractional units 
(see Appendix~\ref{sect:pds_norm}). The $s$ value was computed by 
comparing periodograms at frequencies below $2 \times
10^{-3}$~Hz (above which the Poisson noise power becomes comparable to
the source variability). For each pair of periodograms the value of $s$ was
within the range expected for stationary data (specifically $\left| s
\right| <1$, within one standard deviation of the expected value).

\subsection{rms spectrum}

\begin{figure}
\begin{center}
   \includegraphics[width=6.0 cm, angle=270]{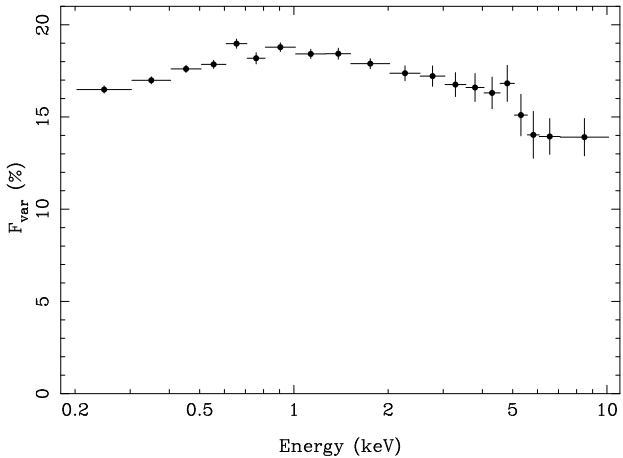}
\end{center}
\caption{
rms spectrum of \mrk\ measured using EPIC pn light curves with 1000~s bins.
\label{fig:mrk766_2}
}
\end{figure}

The variability amplitude as a function of energy was calculated by
measuring $F_{\rm var}$ from light curves extracted in various energy
ranges. The results are shown in Fig.~\ref{fig:mrk766_2} and 
the errors were calculated
using equation~\ref{eqn:fvar_err1} to account for the effect of
Poisson noise. 
The variability amplitude is clearly a function of energy, i.e. \mrk\
shows significant spectral variability. This was confirmed by a
Fourier analysis of the light curves in different energy bands (Vaughan \&
Fabian 2003) which revealed complex energy-dependent variability.

The rms spectrum was re-calculated for light curves
containing only single pixel (pattern 0) events and again for double
pixel (patterns 1--4) events.  These two sets of data were extracted
from the  same detector and using identical extraction regions etc.  After
accounting for the difference in count rate between  single and double
pixel events the two sets of light curves should be identical except
for the effects of Poisson noise. The two rms spectra should  be the
same except for  Poisson errors.
Comparing the ratio of the two rms spectra using a $\chi^2$-test
(against the hypothesis of unity ratio) gave  $\chi^2=25.3/20$ degrees
of freedom.  Comparing the difference of
the two rms spectra to the hypothesis of zero difference gave
identical results and shows the two rms spectra are indeed fairly
consistent.  
This test indicates that for real 
data the error formula given above does provide a reasonable
description of the uncertainty induced by photon noise.

\section{Discussion}
\label{sect:disco}

The analysis of stochastic processes, such as X-ray variability  of
AGN, is conceptually different from the analysis of deterministic data
such as time-averaged spectra (see discussions in e.g. Jenkins \&
Watts 1968; Priestley 1981; Bendat \& Piersol 1986; Bloomfield 2000).
For example, when observing the  spectrum of a constant source one
expects repeatability of the data to within the limits set by
measurement errors, i.e. each new realisation of the spectrum should
be consistent within the errors.  In AGN variability analysis it is
the signal itself that is randomly variable; one does not expected
repeatability of  quantities such as the mean or variance.  These
statistical moments will change (randomly) with each new  light curve
even if there are no measurement errors.

The stochastic nature of red noise processes means that it is usually
only their {\it average} properties that can provide physical insight.
Non-deterministic data should be handled statistically.  For example,
it is customary to examine the timing properties of X-ray binaries
using PSDs estimated from the average periodogram of an
ensemble of light curves (e.g. van der Klis 1995).  Averaging over
many independent realisations reduces the random fluctuations inherent
in the noise process.

In most AGN timing studies however there are rarely enough data to
construct averages in this way (but see Papadakis \& Lawrence 1993 and
Uttley \et 2002 for more on PSD estimation for AGN). As a result of
this relative lack of data, AGN timing studies often emphasise the
properties of a single light curve. But emphasis on the detailed
properties of any single realisation of a stochastic process can be
misleading.  For example, AGN light curves show large fluctuations in
variance. These changes provide little insight as they are expected
even when the underlying physical process responsible for the
variability is constant.  Rather, they may simply be statistical
fluctuations intrinsic to the stochastic process. All red noise
processes show random fluctuations in both mean and variance and the
variance will be distributed in a non-Gaussian fashion with a large
scatter (see sections~\ref{sect:stochasticity} and
\ref{sect:errors_intrinsic}).

Previous claims of non-stationary variability based on changes in
variance (e.g. Nandra \et 1997; Dewangan \et 2002; Gliozzi, Sambruna
\& Eracleous 2003) should therefore be treated with caution  since
they did not account for this intrinsic scatter (see also section
3.3.1 of Leighly 1999 for a discussion of this point).  Real changes
in the PSD would indicate genuine non-stationarity and reflect real
changes in the physical conditions of the variability process. Such
changes can be measured from the average properties of the light
curve, such as the averaged periodogram or the averaged variance (see
section~\ref{sect:errors_intrinsic}).

A different issue is that differences between the variance of 
simultaneous light curves obtained in different energy bands
can be examined using the excess variance (or $F_{\rm var}$) statistic.
It is possible to
estimate the uncertainty in the excess variance due to errors in the
flux measurements. This uncertainty, accounting only for measurement
(e.g. Poisson) errors, can be used when testing for spectral variability,
as demonstrated in section~\ref{sect:mrk766}.

Estimators such as the excess variance provide a useful, if crude,
means of quantifying the variability of AGN. Even  though the
stochastic nature of AGN light curves makes it difficult to estimate
variability amplitudes robustly from short observations, the excess
variance can provide useful  information. For example an analysis of
the excess variances measured from short observations of Seyfert 1
galaxies demonstrated that the variability amplitude (at a given range
of timescales) is inversely correlated with the luminosity of the
source (Nandra \et 1997; Leighly 1999; Markowitz \& Edelson
2001). Although random fluctuations in  variance are expected for AGN
light curves the range of variances observed is far larger than could
be accounted for by this effect alone.  Another example is given in
section~\ref{sect:mrk766_stationary} when it is demonstrated that the
average variance of \mrk\ is a function of the flux of the source. A
similar effect has been observed in  X-ray binaries (Uttley \& M$^{\rm
c}$Hardy 2001).  A discussion of the implications of this result will
be given in Uttley \et in prep.
 
\section{Conclusions}
\label{sect:conc}

This paper discusses some aspects of quantifying the variability
of AGN using simple statistics such as the variance. Various possible
uses of these are presented and some possible problems with their
significance and interpretation are brought to light. The primary
issues are as follows:

\begin{enumerate}

\item 
In order to search for non-stationary variability in an ensemble of
short light curves (or short light curve segments) one can test
whether the individual variances are 
consistent with their mean. Two practical methods are presented
(sections~\ref{sect:bin_var} and \ref{sect:sim_table}). 

\item
In the first method 
the mean variance and its error are calculated at various epochs
by binning the individual variance estimates. This is most useful 
when searching for subtle changes in variability amplitude but
requires large datasets (in order that the variance can be
sufficiently averaged).

\item
In the second method
the individual variance estimates are compared with the expected
scatter around the mean. The expected scatter is calculated using
Monte Carlo simulations of stationary 
processes. The table gives some examples of the scatter expected
for various PSD shapes typical of AGN. This table can therefore be
used to provide a `quick look' at whether the observed 
fluctuations in the variance are larger than expected.
One drawback is that, because the intrinsic scatter 
in the variance is rather large for red noise data, this method is
only sensitive to very large changes in the variability amplitude.
Another drawback is that one has to assume a shape for the PSD.

\item
The excess variance can also be used to quantify how the
variance changes as a function of energy
(section~\ref{sect:spec_var}). An approximate
formula is presented (based on the results of Monte Carlo simulations)
that gives the expected error in the excess variance resulting 
from only observation uncertainties (flux errors such as Poisson noise).
This can be used to test for significant differences in variance
between energy bands. If the normalised excess variances (or $F_{\rm
  var}$s) are found to differ significantly between energy bands this
implies the PSD is energy dependent and/or there are independently
varying spectral components. 

\item
Possibly the most robust yet practical approach to variability
analysis from AGN data is to test the validity of hypotheses
using Monte Carlo simulations. This approach has yielded reliable
PSD estimates for Seyfert galaxies (Green \et 1999; Uttley \et 2002; 
Vaughan \et 2003; Markowitz \et 2003) and has been used to test
the reliability of cross-correlation results (e.g. Welsh 1999) amongst
other things. Section~\ref{sect:simulations} discusses some methods
for simulating red noise data.

\end{enumerate}

\section*{Acknowledgements}

We are very grateful to the referee, Andy Lawrence, for a thoughtful
report that prompted significant improvements to the manuscript.  SV
and PU acknowledge support from PPARC.  This paper made use of
observations obtained with \xmm, an ESA science mission with
instruments and contributions directly funded by ESA Member States and
the USA (NASA).


\appendix

\section{Periodogram normalisation}
\label{sect:pds_norm}

The periodogram is calculated by normalising the modulus-squared of
the DFT (see equation~\ref{eqn:ft}):

\begin{equation}
P(f_{j}) = A |DFT(f_{j})|^{2}.
\end{equation}

There are a variety of options for the normalisation $A$ used in the
literature, each has desirable properties. In the mathematical
literature on time series analysis a normalisation of the form $A=2/N$
is standard (e.g. Priestley 1981; Bloomfield 2000). However, this
normalisation is generally not used for time series analysis in astronomy
because the periodogram then depends on flux of the source and the
binning of the time series. 
Below are listed three of the most commonly used normalisations,
which only differ by factors of $\bar{x}$, the mean count rate in cts/s
($A_{\rm abs}=\bar{x} A_{\rm Leahy} = \bar{x}^2
A_{\rm rms^{2}}$).
The factor of two is present in all these normalisations to make the periodogram
`one sided,' meaning that integrating over positive
frequencies only yields the correct variance.

\begin{enumerate}

\item
$A_{\rm rms^{2}}=2 \Delta T_{\rm samp}/\bar{x}^{2} N$ -- defined by
van der Klis (1997) 
(see also Miyamoto \et 1991). This is the 
normalisation most often used in analysis of AGN and X-ray
binaries because the integrated periodogram yields the fractional
variance of the data. The units for the
periodogram ordinate are (rms/mean)$^{2}$ Hz$^{-1}$ (where rms/mean is
the dimensionless quantity $F_{\rm{var}}$), or simply Hz$^{-1}$.

If a light curve consists of a binned photon counting signal
(and in the absence of other effects such
as detector dead-time)  the expected Poisson noise `background'
level in its periodogram is given by  
\begin{equation}
P_{\rm noise} = \frac{2(\bar{x}+B)}{\bar{x}^{2}} 
\frac{\Delta T_{\rm samp}}{\Delta T_{\rm bin}},
\end{equation}
where $\bar{x}$ is the mean source count rate, $B$ is the mean
background count rate, $\Delta T_{\rm samp}$ is the sampling interval
and $\Delta T_{\rm bin}$ is the time bin width. 
The factor of $ \Delta T_{\rm samp} / \Delta T_{\rm bin}$ accounts for 
aliasing of the Poisson noise level if the original photon counting
signal contained gaps. If the light curve is
a series of contiguous time bins (i.e. $\Delta T_{\rm bin} = \Delta T_{\rm
samp}$) and has zero background (which is approximately true for many
\xmm\ light curves of AGN) then this reduces to $P_ {\rm noise} =
2/\bar{x}$. 

For a light curve with Gaussian errors $\sigma_{{\rm err}, i}$ the
noise level in the periodogram is
\begin{equation}
P_{\rm noise} = 
\frac{2 \Delta T_{\rm bin} \overline{\sigma_{\rm{err}}^{2}}}{\bar{x}^{2}}
\frac{\Delta T_{\rm samp}}{\Delta T_{\rm bin}}.
\end{equation}

\item
$A_{\rm Leahy}=2 \Delta T_{\rm samp}/\bar{x}N$ -- originally due to
Leahy \et (1983). This has the
property that the expected Poisson noise level is simply 2 (for continuous,
binned photon counting data). If the 
light curve consists only of Poisson fluctuations then the periodogram
should be distributed exactly as $\chi_{2}^{2}$. 
It is this property that makes this normalisation the standard for
searching for periodic signals in the presence of Poisson noise (see
Leahy \et 1983).
If the input light curve is in units of ct s$^{-1}$ then the
periodogram ordinate is in units of ct s$^{-1}$ Hz$^{-1}$. 

\item
$A_{\rm abs}=2 \Delta T_{\rm samp}/N$ -- this is the normalisation used in
equation~\ref{eqn:pds}. 
This gives the periodogram in absolute units [e.g. (ct
s$^{-1}$)$^{2}$ Hz$^{-1}$] and so the integrated periodogram gives the total
variance in absolute units [e.g. (ct s$^{-1}$)$^{2}$]
For a contiguously binned light curve with Poisson errors
the noise level is $P_{\rm noise} = 2\bar{x}$, and for
Gaussian errors the noise level is
$P_{\rm noise} = 2 \Delta T_{\rm bin} \overline{\sigma_{\rm{err}}^{2}}$.

\end{enumerate}

%

\section{Monte Carlo demonstration of Poisson noise induced
  uncertainty on excess variance} 
\label{sect:MC}

To estimate the effect on $\sigma_{\rm{NXS}}^2$ due only to Poisson
noise the basic strategy was as follows.  

\begin{enumerate}

\item Generate a random red noise light curve. This acts as the
`true' light curve of the source.

\item Add Poisson noise, i.e. draw fluxes from the light curve according to the Poisson
distribution. This simulates `observing' the true light curve. Error
bars were assigned based on the `observed' counts in each bin ($\sqrt{counts}$). 

\item Measure the normalised excess variance
$\sigma_{\rm{NXS}}^2$ of the observed light curve.
This will be different from the variance of the true light curve
because of the Poisson noise.

\end{enumerate}

\begin{figure}
\begin{center}
   \includegraphics[width=5.0 cm, angle=270]{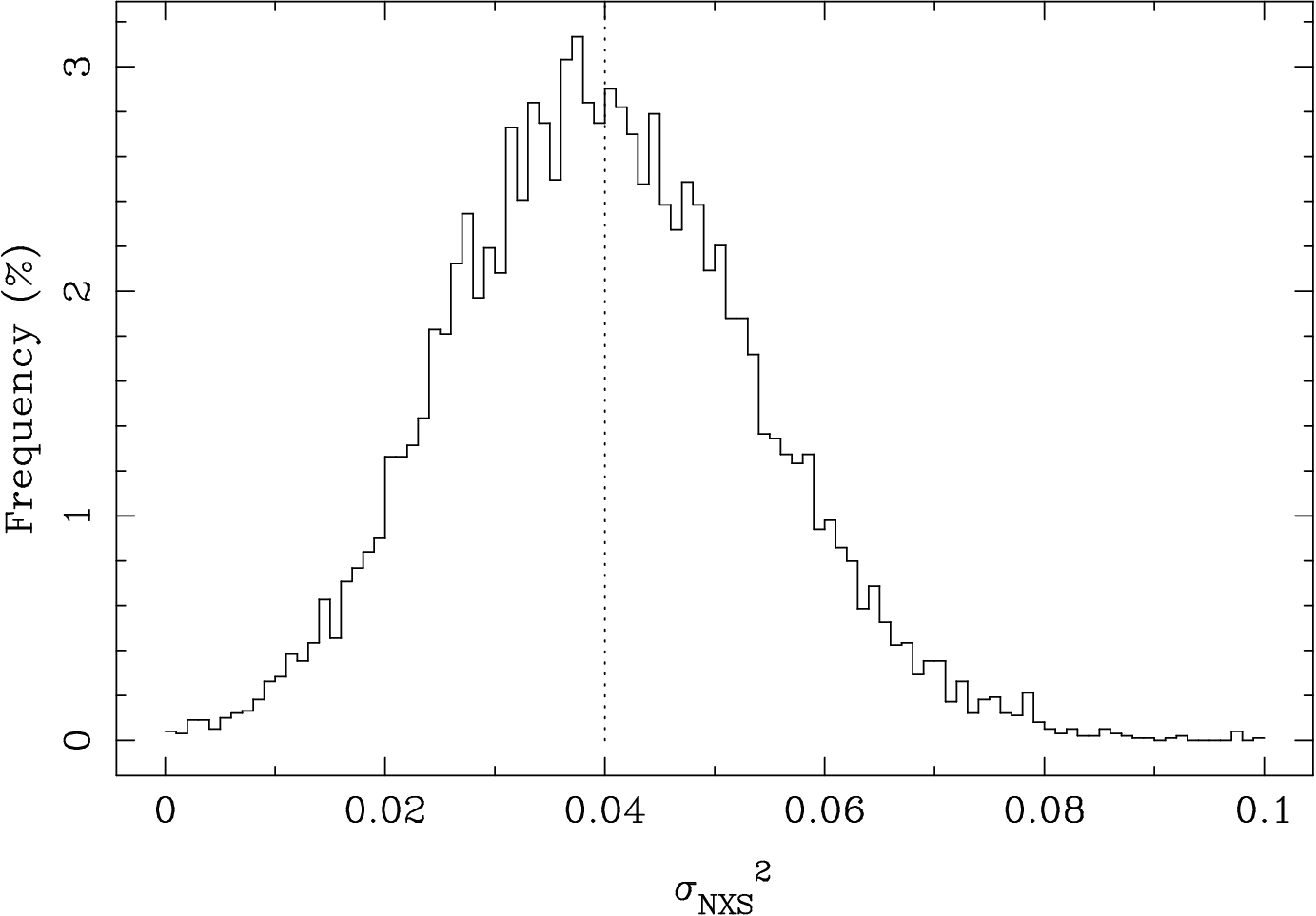}\\
\vspace{0.2 cm}
   \includegraphics[width=5.0 cm, angle=270]{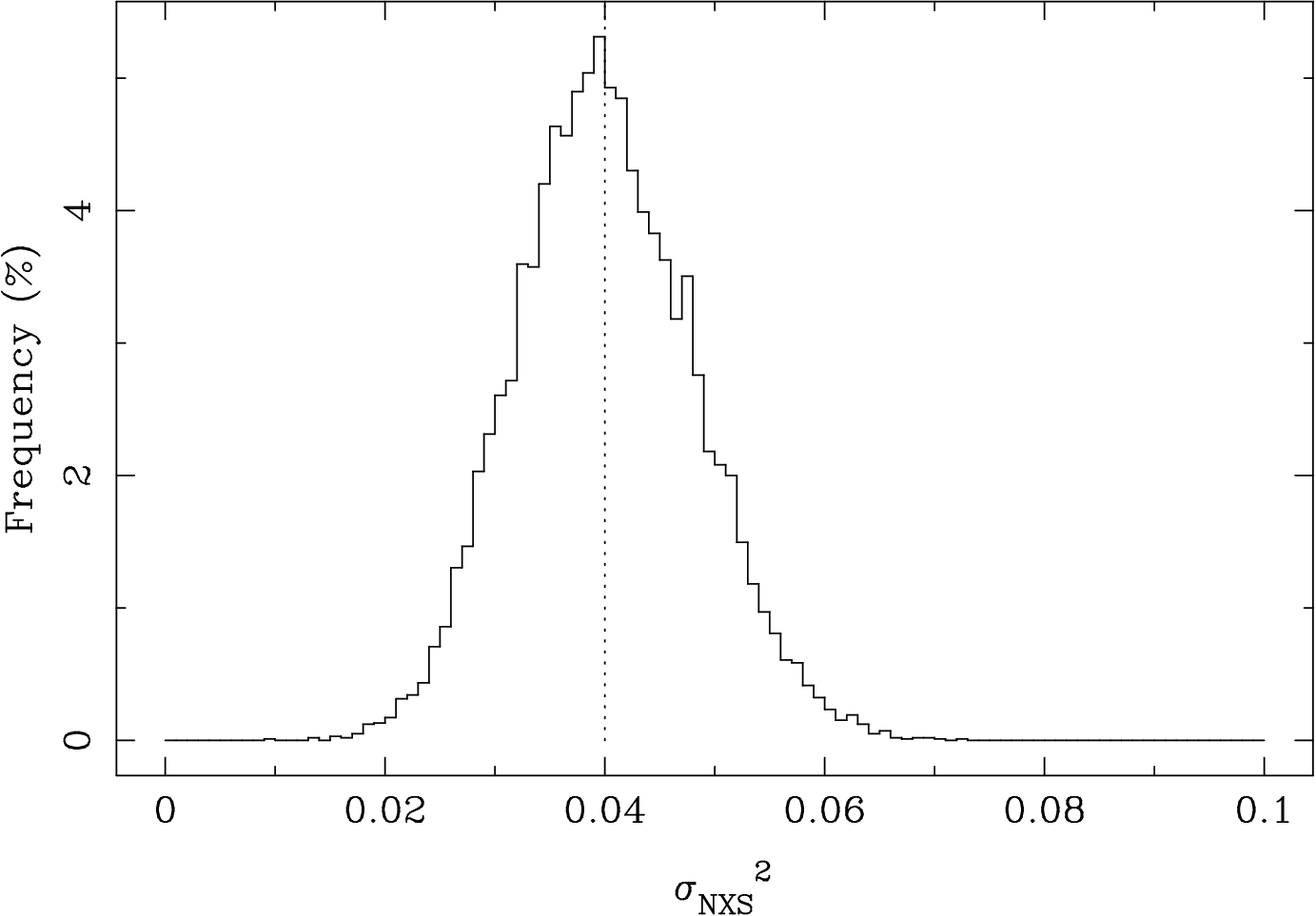}\\
\vspace{0.2 cm}
   \includegraphics[width=5.0 cm, angle=270]{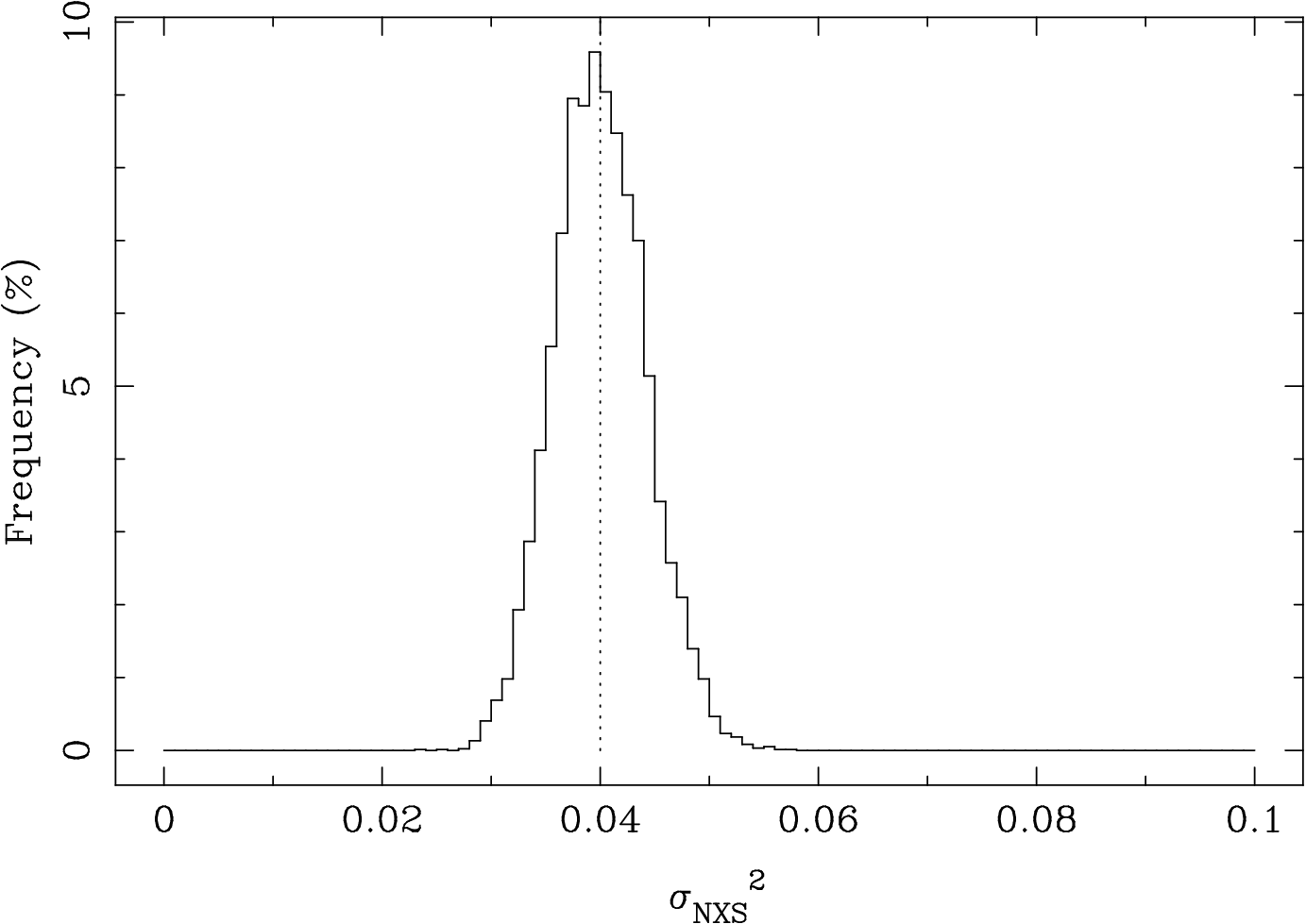}
\end{center}
\caption{
Distribution of measured $\sigma_{\rm{NXS}}^2$ from 10,000
`observations' of the same light curve. In each case the `true'
$\sigma_{\rm{NXS}}^2$ is 0.04
(dotted line).  The top panel used the lowest S/N data,
the bottom panel used the highest S/N data.
The mean number of counts per bin in the simulated light curves
was 15 (top), 30 (middle) and 100 (bottom).
As the S/N increases (count rate increases) the distribution of
$\sigma_{\rm{NXS}}^2$ becomes narrower. 
(Note: this is different
from Fig.~\ref{fig:fvar_dist1}, which shows how the variance changes
between different realisations of the same stochastic process.)
\label{fig:fvar_dist2}
}
\end{figure}

Steps 2 and 3 were repeated, using the same true light curve, to
obtain the distribution of $\sigma_{\rm{NXS}}^2$.
Fig.~\ref{fig:fvar_dist2} shows some results.  In this example the
`true' light curve was generated with a $f^{-2}$ PSD and normalised
to a pre-defined mean and variance, e.g. $S^{2}/\bar{x}^{2} = 0.04$
($F_{\rm{var}}=20$\%).  This light curve was then observed (i.e. steps
2 and 3 were repeated) $10^4$ times\footnote{As this measures only the effect
due to Poisson noise, the results are largely independent of the
details of the light curve, including the PSD, as long as the flux is
non-zero throughout the light curve.  This was confirmed by repeating
the  above experiment using data produced from PSD slopes in the range
$\alpha=0-2$.}.  The three panels correspond to different mean count
rates for the  true light curve (i.e. different S/N of the
observation).  The ($1\sigma$) widths of the $\sigma_{\rm{NXS}}^2$
distributions are Monte-Carlo estimates of the size of the error bars
on $\sigma_{\rm{NXS}}^2$ due to Poisson noise.

As is clear from Fig.~\ref{fig:fvar_dist2} the distribution of
$\sigma_{\rm{NXS}}^2$ becomes narrower, i.e. the error on
$\sigma_{\rm{NXS}}^2$ gets smaller, as the S/N of the data increases.
Obviously in the limit of very high S/N data the measured value of
$\sigma_{\rm{NXS}}^2$ will tend to the `true' value (in this case $0.04$),
i.e. $err(\sigma_{\rm{NXS}}^2) \rightarrow 0$ as $counts \rightarrow
\infty$.  It should also be noted that
the distributions are quite symmetrically centred on the correct value, indicating that
$\sigma_{\rm{NXS}}^2$ is an unbiased estimator of the intrinsic
variance in the light curve, even in relatively low S/N data.

\begin{figure}
\begin{center}
   \includegraphics[width=6.0 cm, angle=270]{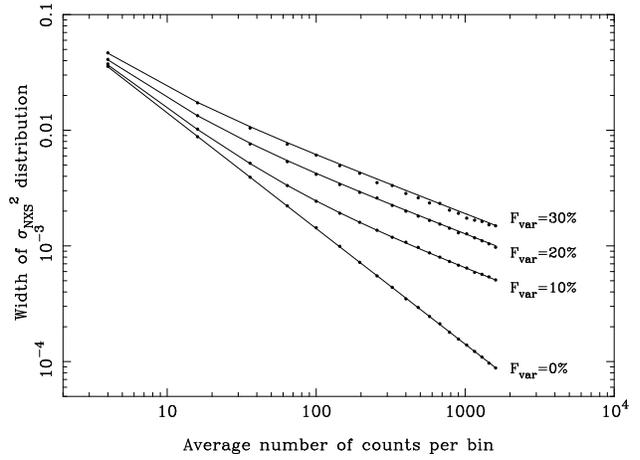}
\end{center}
\caption{
Width of the distribution of $\sigma_{\rm{NXS}}^2$ (resulting from
Poisson noise) as a function of the number of counts per bin.  Compare
with Fig.~\ref{fig:fvar_dist2}.  The solid curve shows the function
described in the text (equation~\ref{eqn:xs_error}).
\label{fig:signal/noise}
}
\end{figure}

In order to assess how the error on $\sigma_{\rm{NXS}}^2$ changes with
S/N, the width of its distribution was measured from simulated data at
various different settings of S/N and intrinsic variance (i.e.
$S^{2}/\bar{x}^{2}$). Width of the distribution at each setting was
calculated from only $500$ `observations' of each light curve. In
order that no particular realisation  adversely affect the outcome,
and to increase the statistics, this was repeated for $20$ different
random light curves (of the same fractional variance) and the width of
the $\sigma_{\rm{NXS}}^2$ distributions were  averaged (i.e. the whole
cycle of steps 1--3 was repeated $20$ times).  Thus for each specified
value of S/N and fractional variance, the error on
$\sigma_{\rm{NXS}}^2$ is estimated from $10^4$ simulated
`observations.'  These Monte Carlo estimated errors on the normalised
excess variance  are shown in Fig.~\ref{fig:signal/noise}.

The solid lines show the functions defined by
equation~\ref{eqn:xs_error} (which was obtained by fitting various
trial functions to the Monte Carlo results).  Clearly this equation
gives a very good match to the Monte Carlo results.

If the variability is not well detected, either because the S/N is low
or the intrinsic amplitude is weak, then $S^2 \approx
\overline{\sigma_{\rm{err}}^{2}}$. It is the first term on the right
hand side of equation~\ref{eqn:xs_error} that dominates.  If the
variability is well detected, i.e.  $S^{2} \gg
\overline{\sigma_{\rm{err}}^{2}}$, then it is the second term that
dominates.

\begin{equation}
err(\sigma_{\rm{NXS}}^{2}) \approx \left\{ \begin{array}
{r@{\quad:\quad}l}
\sqrt{\frac{2}{N}}\cdot\frac{\overline{\sigma_{\rm{err}}^{2}}}{\bar{x}^{2}}
&
S^2 \approx \overline{\sigma_{\rm{err}}^{2}} \vspace{0.2 cm}
\\
\sqrt{\frac{\overline{\sigma_{\rm{err}}^{2}}}{N}} \cdot \frac{ 2 F_{\rm{var}}}{\bar{x}} 
&
S^2 \gg \overline{\sigma_{\rm{err}}^{2}} .
\end{array} \right.
\label{eqn:nxs_err2}
\end{equation}

In the former case the
deviations from the mean are dominated by the errors and the fluxes are
approximately normally distributed. In this regime 
the error equation becomes the same as
that given in equation~A9 of Edelson \et (2002). 
In the latter case 
the deviations in the light curve are enhanced by the intrinsic 
variance. The second term is similar to the first except multiplied
by a factor $\sqrt{2 \sigma_{\rm XS}^{2} / \overline{\sigma_{\rm err}^{2}}}$
to account for this.

Equation~\ref{eqn:xs_error} can be used to give the uncertainty on
$F_{\rm{var}}$ thusly
\begin{eqnarray}
\lefteqn{
err(F_{\rm{var}}) =
\frac{1}{2 F_{\rm{var}}} err(\sigma_{\rm{NXS}}^{2}) = }
\nonumber \\
\lefteqn{\qquad
\sqrt{ \left\{ \sqrt{\frac{1}{2N}} \cdot\frac{ \overline{\sigma_{\rm{err}}^{2}}
}{  \bar{x}^{2}F_{\rm{var}} }  \right\}^{2}
+
\left\{ \sqrt{\frac{\overline{\sigma_{\rm{err}}^{2}}}{N}}
\cdot\frac{1}{\bar{x}}  \right\}^{2}  },
}
\label{eqn:fvar_err1}
\end{eqnarray}
and this is the equation used to derive the errors shown in Fig.~\ref{fig:mrk766_2}.
In the two regimes this becomes:
\begin{equation}
err(F_{\rm{var}}) \approx \left\{ \begin{array}
{r@{\quad:\quad}l}
\sqrt{\frac{1}{2N}} \cdot\frac{ \overline{\sigma_{\rm{err}}^{2}}  }{
\bar{x}^{2}F_{\rm{var}} }&
S^2 \approx \overline{\sigma_{\rm{err}}^{2}}\vspace{0.2 cm}
\\
\sqrt{\frac{\overline{\sigma_{\rm{err}}^{2}}}{N}} \cdot\frac{1}{\bar{x}} &
S^2 \gg \overline{\sigma_{\rm{err}}^{2}}.
\end{array} \right.
\end{equation}
In the first instance, when the variability is not well
detected,
$\sigma_{\rm{NXS}}^{2}$ should be preferred over $F_{\rm{var}}$
as negative values of $\sigma_{\rm{NXS}}^{2}$ are possible.
Additional Monte Carlo simulations confirmed the above equations are
valid for both Gaussian and Poisson distributed flux errors.
It is worth reiterating that this error accounts only for
measurement errors on the fluxes. It does not account for the
intrinsic scatter in the fluxes inherent in any red noise
process.


\bsp
\label{lastpage}
\end{document}